\documentclass{aastex63}

\usepackage{xcolor}
\usepackage{subfigure}

\shorttitle{\textit{FOXSI-2} Solar Microflares II}
\shortauthors{Vievering et al.}

\begin{document}

\title{\textit{FOXSI-2} Solar Microflares II: Hard X-ray Imaging Spectroscopy and Flare Energetics}

\correspondingauthor{Juliana T. Vievering}
\email{vieve004@umn.edu}

\author{Juliana T. Vievering}
\affil{University of Minnesota, Twin Cities, Minneapolis, MN, USA}

\author{Lindsay Glesener}
\affil{University of Minnesota, Twin Cities, Minneapolis, MN, USA}

\author{P. S. Athiray}
\affil{University of Minnesota, Twin Cities, Minneapolis, MN, USA}
\affil{NASA Postdoctoral Program, NASA Marshall Space Flight Center, Huntsville, AL, USA}

\author{Juan Camilo Buitrago-Casas}
\affil{Space Sciences Laboratory, University of California at Berkeley, Berkeley, CA, USA}

\author{Sophie Musset}
\affil{University of Minnesota, Twin Cities, Minneapolis, MN, USA}
\affil{SUPA, School of Physics \& Astronomy, University of Glasgow, Glasgow G12 8QQ, UK}

\author{Daniel Ryan}
\affil{NASA Goddard Space Flight Center, Greenbelt, MD, USA}

\author{Shin-nosuke Ishikawa}
\affil{Rikkyo University, Graduate School of Artificial Intelligence and Science, Toshima, Tokyo, Japan}

\author{Jessie Duncan}
\affil{University of Minnesota, Twin Cities, Minneapolis, MN, USA}

\author{Steven Christe}
\affil{NASA Goddard Space Flight Center, Greenbelt, MD, USA}

\author{S\"{a}m Krucker}
\affil{Space Sciences Laboratory, University of California at Berkeley, Berkeley, CA, USA}
\affil{University of Applied Sciences and Arts Northwestern Switzerland, Windisch, Switzerland}

\begin{abstract}

We study the nature of energy release and transfer for two sub-A class solar microflares observed during the second flight of the \textit{Focusing Optics X-ray Solar Imager} (\textit{FOXSI-2}) sounding rocket experiment on 2014 December 11. \textit{FOXSI} is the first solar-dedicated instrument to utilize focusing optics to image the Sun in the hard X-ray (HXR) regime, sensitive to the energy range 4-20 keV. Through spectral analysis of the two microflares using an optically thin isothermal plasma model, we find evidence for plasma heated to temperatures of ${\sim}10$~MK and emissions measures down to ${\sim}10^{44}$~cm$^{-3}$. Though nonthermal emission was not detected for the \textit{FOXSI-2} microflares, a study of the parameter space for possible hidden nonthermal components shows that there could be enough energy in nonthermal electrons to account for the thermal energy in microflare 1, indicating that this flare is plausibly consistent with the standard thick-target model. With a solar-optimized design and improvements in HXR focusing optics, \textit{FOXSI-2} offers approximately five times greater sensitivity at 10~keV than the \textit{Nuclear Spectroscopic Telescope Array} (\textit{NuSTAR}) for typical microflare observations and allows for the first direct imaging spectroscopy of solar HXRs with an angular resolution at scales relevant for microflares. Harnessing these improved capabilities to study the evolution of small-scale events, we find evidence for spatial and temporal complexity during a sub-A class flare. These studies in combination with contemporanous observations by the Atmospheric Imaging Assembly onboard the \textit{Solar Dynamics Observatory} (\textit{SDO}/AIA) indicate that the evolution of these small microflares is more similar to that of large flares than to the single burst of energy expected for a nanoflare.

\end{abstract}

\section{Introduction} \label{sec:intro}

In the standard model for solar flares, flares are driven by magnetic reconnection in the corona, at which point a large portion (${\sim}20$-$40\%$) of released magnetic energy is converted to kinetic energy of particles \citep{emslie2012,aschw2016}. Some of these particles then travel toward the solar surface, guided by magnetic field lines, and produce nonthermal bremsstrahlung emission in hard X-rays (HXRs) through interactions with dense chromospheric plasma \citep{brown1971}. This interaction subsequently heats the ambient plasma which then fills the flare loop through the process of chromospheric evaporation and produces thermal bremsstrahlung emission in X-rays \citep{neupert1968}. The X-ray regime is thus important to probe in order to better understand energy release and transport during a flaring event. \\
\indent Current X-ray instrumentation has allowed for the study of a broad range of solar eruptive events from \textit{GOES} (\textit{Geostationary Observational Environmental Satellite}) class A to X20, yet we know that events of this magnitude and frequency cannot produce enough energy to heat the solar corona to observed temperatures \citep{hudson1991}. To address this discrepancy in energy, one proposed theory is that small-scale energy releases called nanoflares occur ubiquitously on the solar surface \citep{parker1988,klimchuk2006}. Above the energy scale of nanoflares, there is a class of small-scale solar flares called microflares that are thought to be similar in structure to large solar flares originating in active regions (ARs), just scaled down in magnitude \citep{hudson1991}. As X-ray instrumentation improves, we can begin to probe the structure and dynamics of these solar microflares to better understand their energy release and their contribution to coronal heating. \\
\indent For a number of years (2002-2018), the \textit{Reuven Ramaty High-Energy Solar Spectroscopic Imager} (\textit{\textit{RHESSI}}) was the state-of-the-art instrument for observing the Sun in HXRs \citep{lin2002}. With the limited sensitivity of its indirect imaging technique, \textit{\textit{RHESSI}} was best used for studying large flares, but also observed a number of microflares ($>$25,000) that were analyzed in a statistical study by \cite{hannah2008} and \cite{christe2008}. Though this study investigates microflares of thermal energies ranging from ${\sim}10^{26}{-}10^{30}$~erg, the construction of a flare frequency distribution with this data set reveals that the limited sensitivity of \textit{\textit{RHESSI}} and the choices made in the study for automated analysis result in missing a large portion of events below ${\sim}10^{28}$~erg. \\
\indent One way to achieve improved sensitivity for observing small-scale solar events below this threshold is by instead utilizing a direct imaging technique. The \textit{Nuclear Spectroscopic Telescope Array} (\textit{\textit{NuSTAR}}), launched in 2012, is the first satellite to use focusing optics for observations in the HXR regime, with an energy range of 3-79~keV \citep{harrison2013}. Though \textit{\textit{NuSTAR}} was originally designed for astrophysical purposes, it has also recently been used for solar HXR observations. From the \textit{\textit{NuSTAR}} solar campaigns, a number of small-scale phenomena in HXRs have been detected and studied, including active regions, microflares, and even quiet Sun flares \citep[e.g.][]{grefen2016,hannah2016,glesener2017,wright2017,kuhar2018,hannah2019,glesener2020,cooper2020}. However, \textit{NuSTAR} was not optimized for solar observations, and the relatively limited detector throughput results in a very low livetime, even for small-scale solar events, and effectively reduces the \textit{NuSTAR} energy range to approximately 3-10~keV. \\
\indent The \textit{Focusing Optics X-ray Solar Imager} (\textit{FOXSI}) is the first direct HXR imager to be optimized for solar observations. With a direct-imaging technique, \textit{FOXSI} offers vast improvements in sensitivity and imaging dynamic range compared to \textit{RHESSI} and is thus better equipped to study small-scale events. Additionally, \textit{FOXSI} improves upon \textit{NuSTAR} by offering higher spatial resolution (9" compared to 18" FWHM) and increased detector livetime, which leads to approximately five times greater sensitivity for measuring faint emission at ${\sim}10$~keV (see Section \ref{sec:mic_compare} for details). \textit{FOXSI} has been flown on three sounding rocket campaigns to date \citep{glesener2016}, and \textit{FOXSI-1}, launched in 2012, produced the first ever focused HXR image of the Sun \citep{krucker2014}. In this paper and a corresponding paper by \cite{athiray2020}, two microflares observed by \textit{FOXSI} during its second sounding rocket flight (\textit{FOXSI-2}) on 2014 December 11 are studied. \cite{athiray2020} (hence, ``Paper I") performs a differential emission measure (DEM) analysis of these microflares using a novel data set combining observations from the \textit{Solar Dynamics Observatory} Atmospheric Imaging Assembly  (\textit{SDO}/AIA), the \textit{Hinode} X-ray Telescope (XRT), and \textit{FOXSI-2}. This paper (``Paper II") focuses on HXR imaging and spectroscopy of the microflares and studies the flare energetics with estimates of thermal and nonthermal energies. Section \ref{sec:data} describes the \textit{FOXSI-2} instrument and flight observations. Section \ref{sec:analysis} presents the timing, spectral, and imaging analyses performed for the two observed microflares, and Section \ref{sec:discuss} provides a discussion of these results. Finally, Section \ref{sec:summary} gives a summary of the study.

\begin{figure}
\centering
\includegraphics[width=0.75\textwidth]{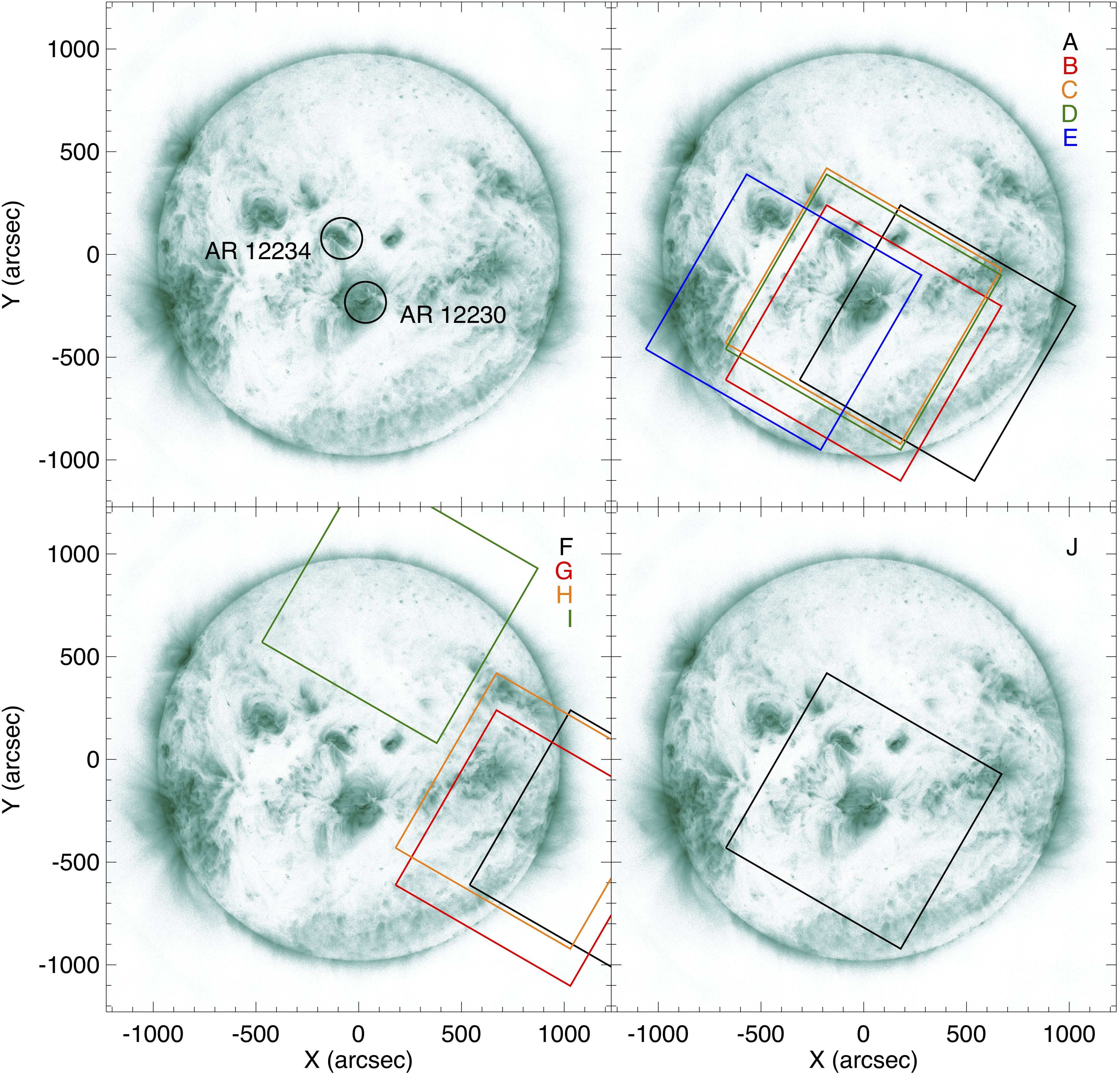}
\caption{AIA $94\mathrm{\AA}$ images during the \textit{FOXSI-2} flight with the flaring active regions identified (circles) and the \textit{FOXSI-2} FOV (16' $\times$ 16') for D6 overlaid (squares). (\textit{Top left}) The \textit{FOXSI-2} microflares occurred in the active regions identified: microflare~1 from AR 12230 and microflare 2 from AR 12234. (\textit{Top right}) Microflare 1 was in the FOV during the first five targets (A-E). A number of pointing adjustments were made early in the flight due to a larger-than-expected observed offset (${\sim}7$') between the experiment and the payload pointing system. (\textit{Bottom left}) \textit{FOXSI-2} observed quiet regions of the Sun during Targets F-I. (\textit{Bottom right}) Microflare 2 was observed during the final target of the flight (J).} 
\label{fig:foxsi_fov}
\end{figure}

\begin{figure}
\centering
\includegraphics[width=0.7\textwidth]{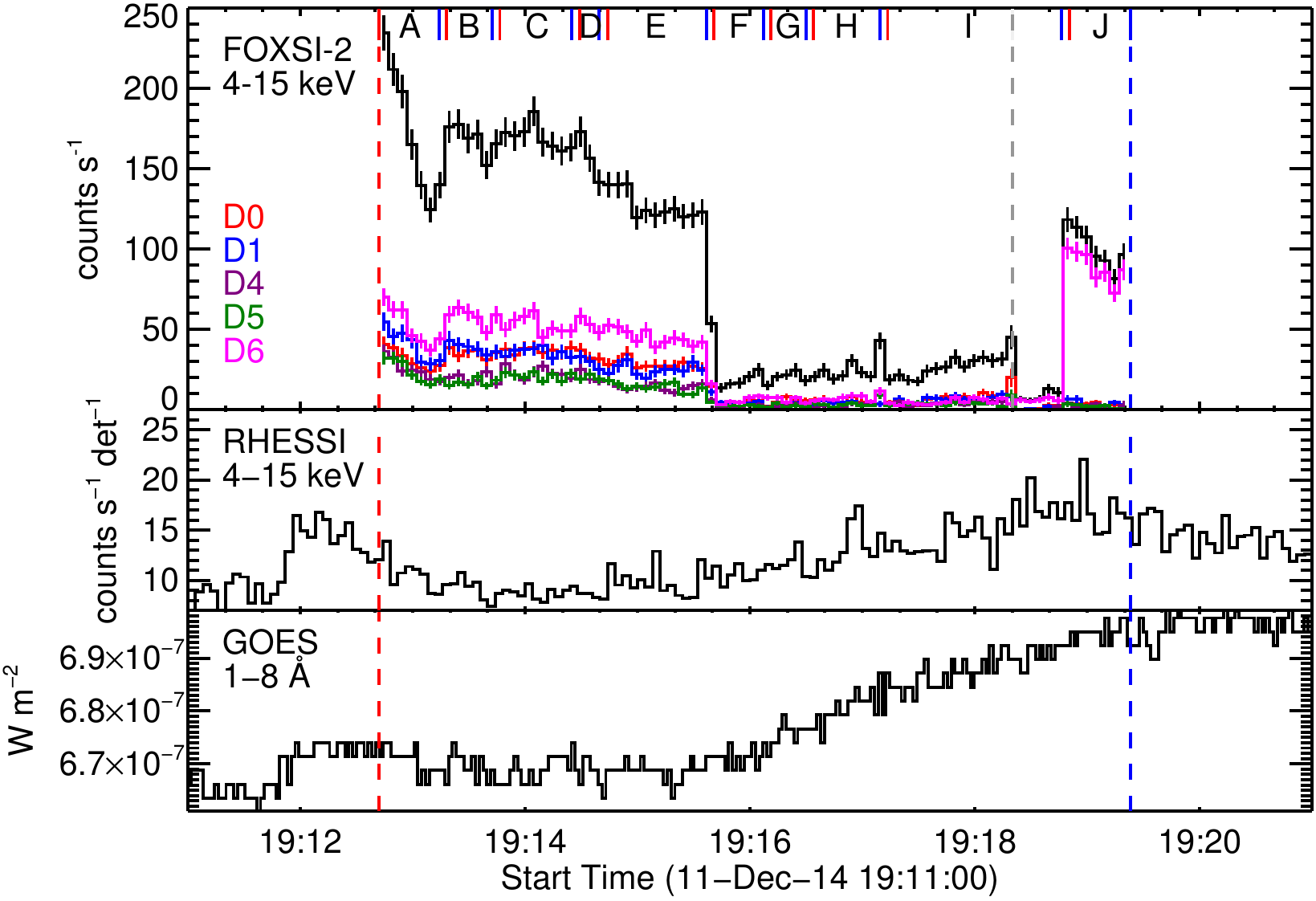}
\caption{(\textit{Top panel}) Light curves of the \textit{FOXSI-2} sounding rocket flight for each Si detector (D0, D1, D4, D5, D6) and for all Si detectors combined over the full FOV (not corrected for vignetting). The observation period lasted ${\sim}6.5$~min, and several targets were observed, which are labeled at the top of the plot (see Table \ref{targets} for microflare targets). The red lines indicate the start of a target (pointing stabilized), and the blue lines indicate the end of a target, with pointing changes taking ${\sim}4$~s. Aluminum attenuators covering 6 of 7 detectors were deployed at 19:18:20 UT (dashed gray line), and after this time, D6 (magenta) was the only detector without an attenuator. The bottom two panels show full Sun light curves from \textit{RHESSI} (4-15~keV) and \textit{GOES} (1.0-8.0~$\mathrm{\AA}$) during the \textit{FOXSI-2} flight.   }
\label{fig:lc_flight}
\end{figure}

\section{\textit{FOXSI-2} Observations and Data} \label{sec:data}

The \textit{FOXSI-2} sounding rocket experiment, optimized for the energy range 4-20~keV, is composed of seven separate telescopes, each made up of an optical module and a semiconductor strip detector. Having multiple telescopes allows for consistency checks between detectors and also provides the opportunity to test a variety of new technologies. Each optical module is developed at Marshall Space Flight Center (MSFC) and is designed as a set of nested shells of the Wolter-1 type with a 2-meter focal length. For \textit{FOXSI-2}, five of the optical modules contained seven mirror shells each while two of the modules were upgraded from the first flight to include ten shells each, increasing the effective area \citep{christe2016}. The semiconductor detectors, developed at the Japan Aerospace Exploration Agency Institute of Space and Astronautical Science (JAXA/ISAS), were made of either silicon (Si; five detectors) or cadmium telluride (CdTe; two detectors). The CdTe detectors were added for the \textit{FOXSI-2} experiment and offer an improvement compared to Si due to having a higher quantum efficiency above 10~keV \citep{ishikawa2016}. For the Si detectors, the field of view (FOV) is ${\sim}16' \times 16'$ while the FOV for the CdTe detectors is ${\sim}13' \times 13'$. The analysis in this paper is focused on data from the Si detectors. \\
\indent The \textit{FOXSI-2} flight took place on 2014 December 11 with an observation period lasting ${\sim}6.5$ minutes. Multiple regions on the Sun were targeted, including a number of active and quiet regions. All targets from the \textit{FOXSI-2} flight are listed in Table \ref{targets} and displayed in Figure \ref{fig:foxsi_fov}. During the flight (see Figure \ref{fig:lc_flight} for light curves), two microflares occurred, one starting just before our observations from AR 12230 (``microflare 1") and one near the end of our observations from AR 12234 (``microflare 2"). When considering other solar X-ray instruments, we note that both microflares were too faint to be flagged for the \textit{GOES} flare catalog; microflare 2 was missed by the \textit{RHESSI} flare catalog, as it did not pass the set threshold for imaging, although \textit{RHESSI} did acquire usable flux for source coalignment (see Section \ref{sec:coalign}). Light curves from \textit{RHESSI} and the \textit{GOES} X-ray Sensor (XRS) during the \textit{FOXSI-2} flight (full Sun) are included in Figure \ref{fig:lc_flight} for comparison. Data from all \textit{FOXSI} flights can be found on the Virtual Solar Observatory, and the data analysis software can be found on the \textit{FOXSI} GitHub page\footnote{https://github.com/foxsi/foxsi-science}. Additional information on data access is available on the \textit{FOXSI} website\footnote{http://foxsi.umn.edu/data}.  

\begin{table}
\begin{tabular}{|c|c|c|c|c|c|c|}
\hline
\textit{Target} & \textit{Center coordinates} & \textit{Start time} & \textit{End time} & \textit{Duration}  & \textit{Flare}  & \textit{Source off-axis angle}  \\ 
 & (arcsec) & (UT) & (UT) & (s) & & (arcmin) \\ \hline
A & {[} 359, -431 {]} & 19:12:42 & 19:13:14 & 32 & \multicolumn{1}{c|}{1} & 6.1\\
B & {[} -1, -431 {]} & 19:13:18 & 19:13:43 & 25 & \multicolumn{1}{c|}{1} & 2.4\\
C & {[} -1, -251 {]} & 19:13:47 & 19:14:25 & 38 & \multicolumn{1}{c|}{1} & 0.4\\
D & {[} -1, -281 {]} & 19:14:29 & 19:14:40 & 11 & \multicolumn{1}{c|}{1} & 0.2\\
E & {[} -390, -281 {]} & 19:14:44 & 19:15:37 & 53 & \multicolumn{1}{c|}{1} & 5.9\\ \hline
F & {[} 1210.5, -431.5 {]} & 19:15:41 & 19:16:07 & 26 & \multicolumn{1}{c|}{...} & ... \\
G & {[} 850, -431.5 {]} & 19:16:11 & 19:16:30 & 19 & \multicolumn{1}{c|}{...} & ... \\
H & {[} 850, -251 {]} & 19:16:34 & 19:17:09 & 35 & \multicolumn{1}{c|}{...} & ... \\
I & {[} 200, 750 {]} & 19:17:14 & 19:18:46 & 92 & \multicolumn{1}{c|}{...} & ... \\ \hline
\multicolumn{7}{|c|}{Attenuator deployed at 19:18:20 UT} \\ \hline
J & {[} 0, -251 {]} & 19:18:51 & 19:19:23 & 32 & \multicolumn{1}{c|}{2} & 5.3 \\ \hline
\end{tabular}
\caption{Target information for the \textit{FOXSI-2} flight. Microflare 1 was observed during Targets A-E (19:12:42$-$19:15:37) and microflare 2 was observed during Target J (19:18:51$-$19:19:23). The source off-axis angle for each flare target is calculated as the distance of the source centroid from the detector center, which is aligned with the optical axis within 1~arcmin. For Targets F-I, \textit{FOXSI-2} was pointed at quiet Sun regions. The provided center coordinates are for the payload pointing system (SPARCS) which differs from the instrument pointing, as discussed in Section \ref{sec:coalign}.  }
\label{targets}
\end{table}

\newpage

\subsection{Raw Image Construction}

Through use of semiconductor strip detectors, we can determine the time, energy, and position of each incoming photon. The detectors are designed as a set of orthogonal strips, with 128 strips on each side. For the Si (CdTe) detectors, the strip pitch is 75$\mu$m (60$\mu$m), corresponding to 7.7" (6.2") on the Sun \citep{ishikawa2011,ishikawa2016,athiray2017}. For each photon event during the flight, the location and signal amplitude for the highest signal strip on each side of the detector is saved, along with that of the two neighboring strips. To produce a raw basic image, the positions of the photons recorded in specified time and energy ranges are first plotted in the detector plane. 
Then, to convert to solar coordinates, the detector image is rotated, translated to the target center, and rebinned according to the new x and y coordinates.

\begin{figure}
\centering
\includegraphics[width=0.75\textwidth]{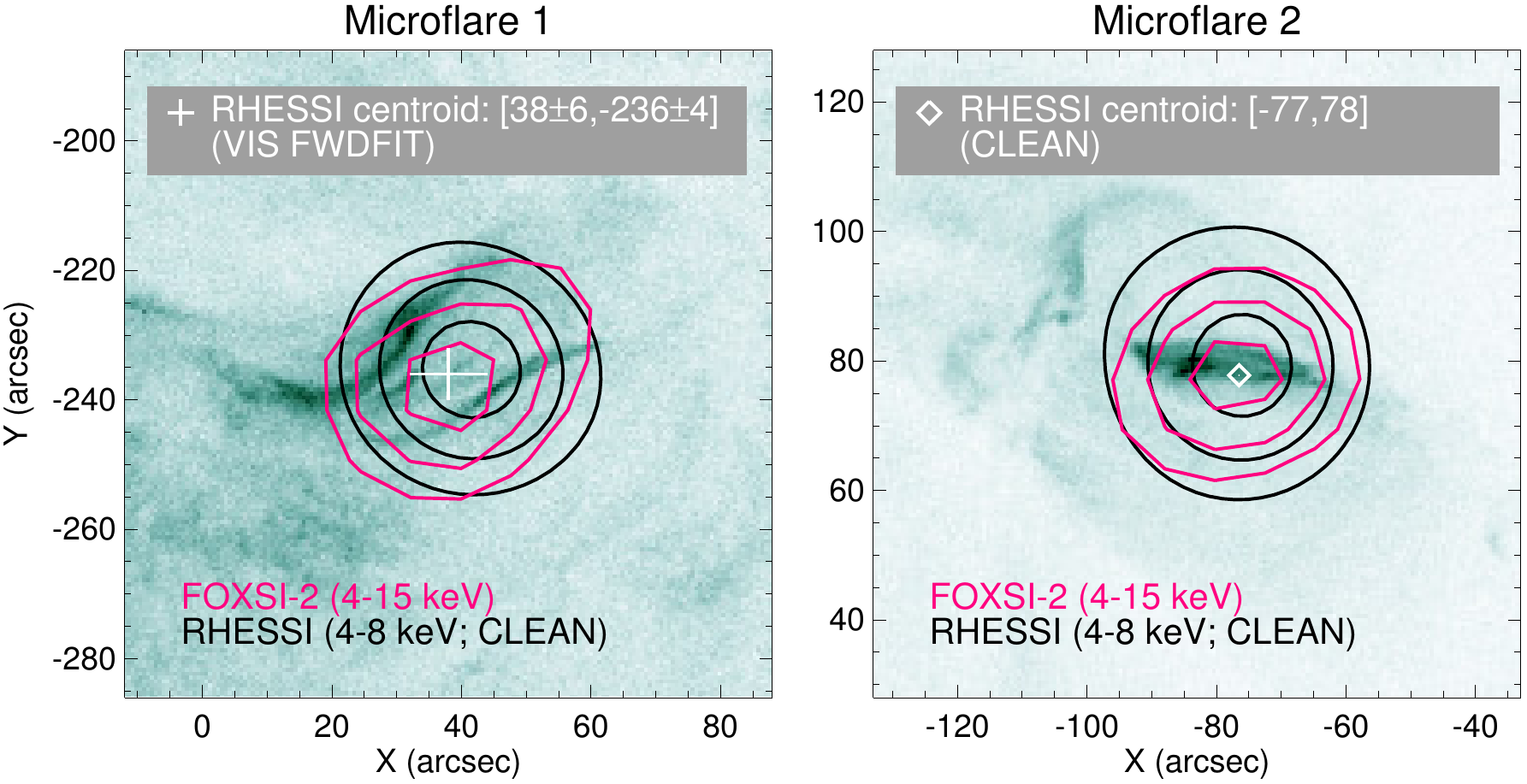}
\caption{Coalignment of \textit{RHESSI} and \textit{FOXSI-2} data for microflares 1 (left) and 2 (right), overlaid on AIA $94\mathrm{\AA}$ images. For microflare 1 (microflare 2), the time interval of 19:11:52-19:13:14 UT (19:18:00-19:20:00 UT) is used for the \textit{RHESSI} data to cover the peak of the flare. The \textit{RHESSI} images shown in contours (black; 50\%, 70\%, 90\%) are created using the CLEAN algorithm, with data from detectors 6, 7, and 8.  For microflare 1, the \textit{FOXSI-2} image data (magenta; 50\%, 70\%, 90\%) from Target A (D6) are aligned such that the centroid matches the \textit{RHESSI} centroid (white +) calculated using the VIS FWDFIT algorithm (circular source). The CLEAN centroid is consistent with the VIS FWDFIT centroid within the uncertainty. For microflare 2, VIS FWDFIT does not produce reasonable results, and we instead calculate the centroid of the CLEAN image (white $\diamond$) for intensities above 50\%.  No coalignment with AIA was performed. }
\label{fig:rhessi}
\end{figure}

\subsection{Coalignment with \textit{RHESSI} Data}\label{sec:coalign}

During the \textit{FOXSI-2} flight, the experiment experienced strong vibration due to a combustion instability in the launch vehicle, which may have been the cause of a larger-than-expected observed offset (${\sim}7'$) between the experiment and the payload pointing system (see Table \ref{targets} for pointing system coordinates). This offset resulted in two substantial pointing adjustments at the beginning of the observations to reach the intended first target (Target C). In order to ensure the accuracy of the spatial coordinates for the \textit{FOXSI-2} targets, we perform a coalignment of the \textit{FOXSI-2} data with contemporaneous \textit{RHESSI} data, as \textit{RHESSI} has a precise pointing knowledge of $<1~$arcsec \citep{lin2002}.  \\ 
\indent Using the VIS FWDFIT algorithm (circular source) with \textit{RHESSI} data from subcollimators\footnote{The high-resolution \textit{RHESSI} subcollimators (1-4) are not used in this case because subcollimators 2 and 4 were unsegmented at the time of the \textit{FOXSI-2} flight, and subcollimators 1 and 3 did not provide useful imaging information, possibly due to multiple faint sources on the Sun.} 6, 7, and 8, the centroid of microflare 1 is calculated to be ($38{\pm}6$,~$-236{\pm}4$) arcsec. We note that we need to use a longer time interval for the \textit{RHESSI} data (${\sim}1.2~$min) in order to reconstruct a clear image of the microflare because, as seen in Figure \ref{fig:lc_flight}, the \textit{RHESSI} flux had almost returned to background levels at the start of the \textit{FOXSI-2} observations. \textit{FOXSI-2}, however, measured ample flux at this time, highlighting the benefits of a direct imaging method for observing small-scale solar events. A correction is then applied to the \textit{FOXSI-2} centroids for each detector and each individual target on microflare 1 such that the \textit{FOXSI-2} centroids match the \textit{RHESSI} centroid coordinates. For microflare 2, the VIS\ FWDFIT algorithm does not produce reasonable results, so we instead calculate the centroid of a \textit{RHESSI} image produced with the CLEAN algorithm. The aligned \textit{RHESSI} and \textit{FOXSI-2} data (D6) for each microflare are shown in Figure \ref{fig:rhessi}.

\section{Analysis} \label{sec:analysis}

\begin{figure}
\centering
\subfigure{
\includegraphics[width=0.63\textwidth]{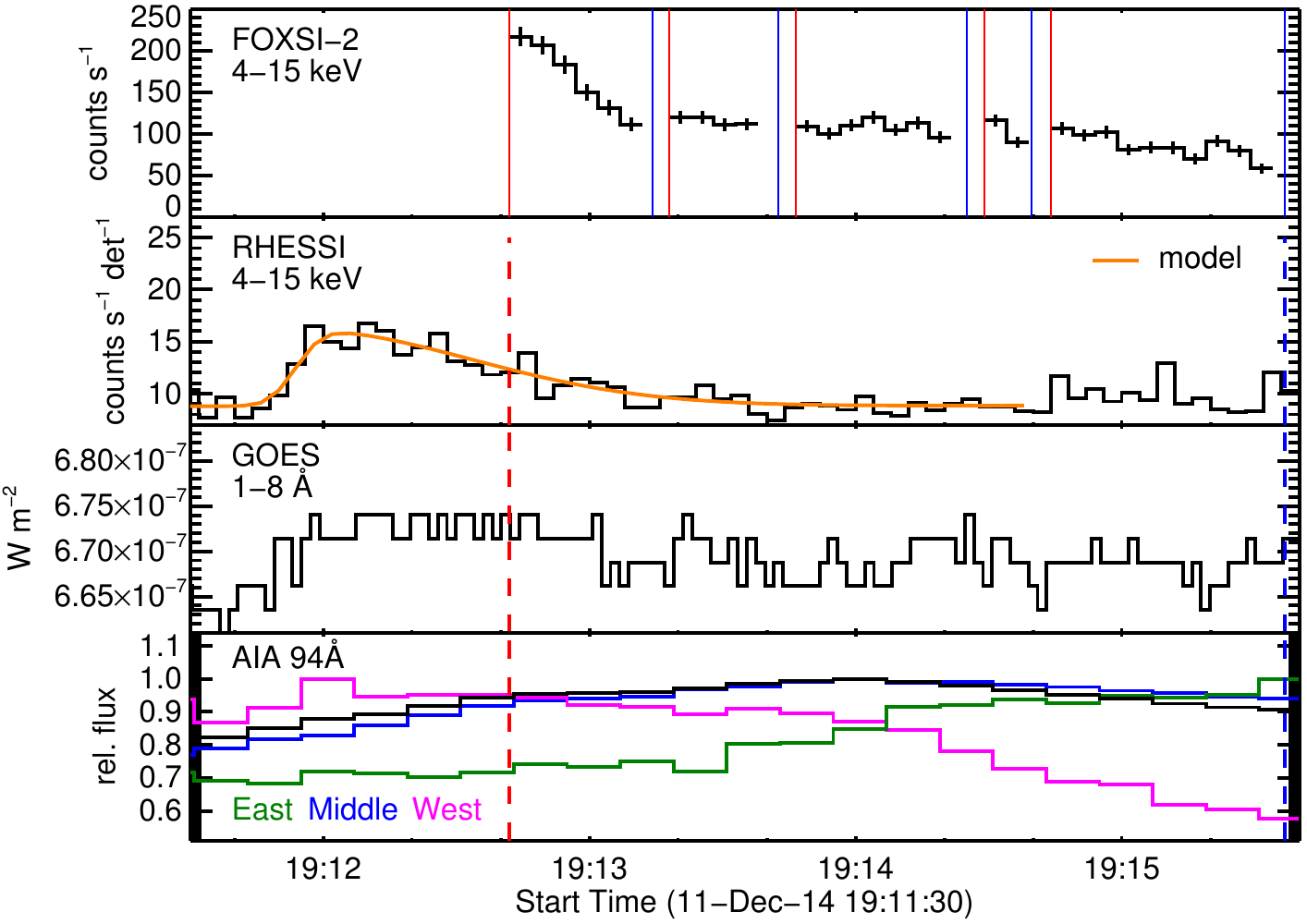}}
\subfigure{
\includegraphics[width=0.35\textwidth]{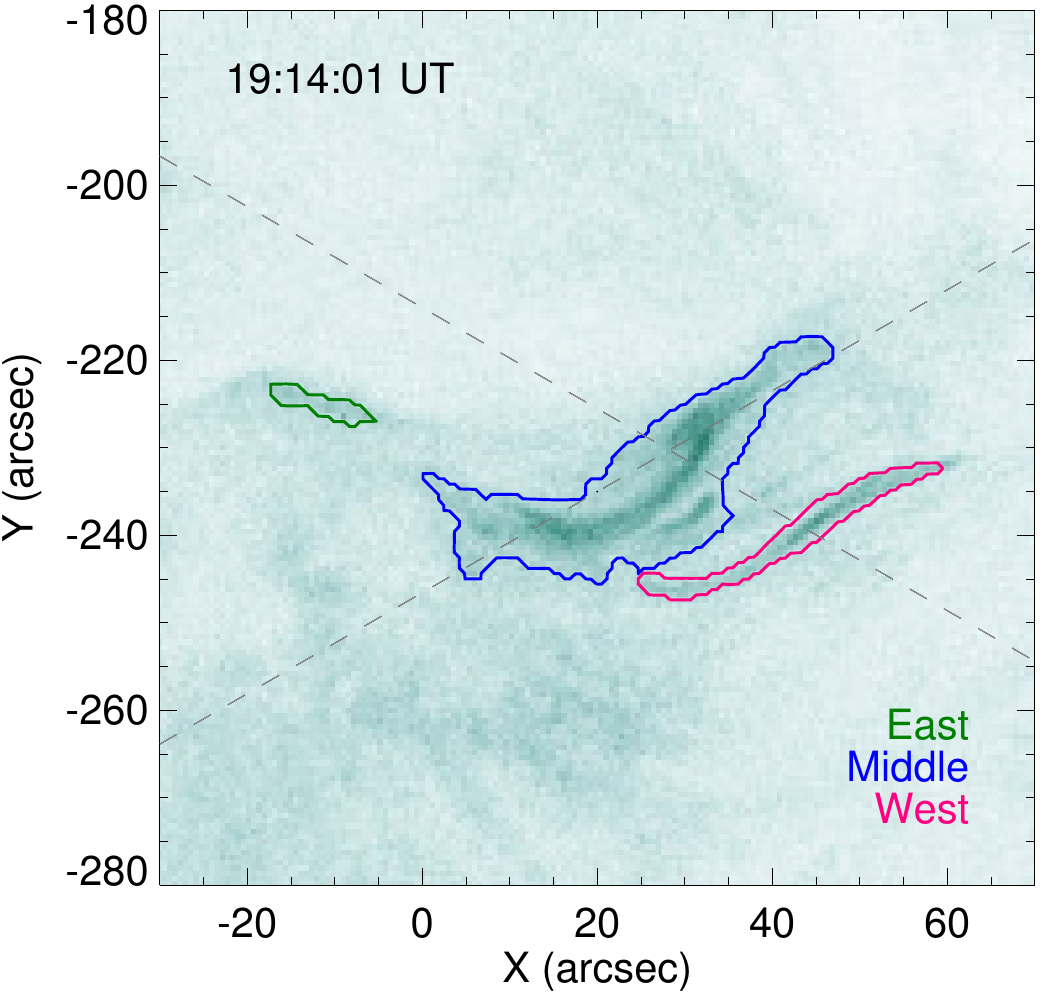}}
\caption{Light curves of microflare 1 from \textit{FOXSI-2}, \textit{RHESSI}, \textit{GOES}, and \textit{SDO}/AIA. This microflare was observed by \textit{FOXSI-2} (top panel) during the first five targets of the flight (Targets A-E). Red vertical lines indicate the start of a target and blue vertical lines indicate the end of a target (${\sim}4$~seconds needed to stabilize pointing). The \textit{FOXSI-2} data show the count rate within a circle (radius 100") centered on microflare 1 for the five Si detectors combined over the energy range 4-15~keV. The count rate for each target is corrected for vignetting effects. The middle two panels show full Sun light curves from \textit{RHESSI} (4-15~keV), with the best fit light curve model (skewed Gaussian + linear background) overlaid, and \textit{GOES} (1.0-8.0~$\mathrm{\AA}$). For the AIA data (94$\mathrm{\AA}$), light curves were extracted from the eastern, middle, and western features highlighted in the image on the right (intensity $>$ $30\%$); the curve for each feature is normalized to its own maximum value within the plotted time range. Flux profiles measured along the dashed lines in the AIA images are presented in Figure \ref{fig:rhessi_intensity}. We note that \textit{FOXSI-2} measured ample flux during the declining phase of this microflare, highlighting the benefits of a direct imaging technique. }
\label{fig:lc_mf1}
\end{figure}

\begin{figure}
\centering
\subfigure{
\includegraphics[width=0.63\textwidth]{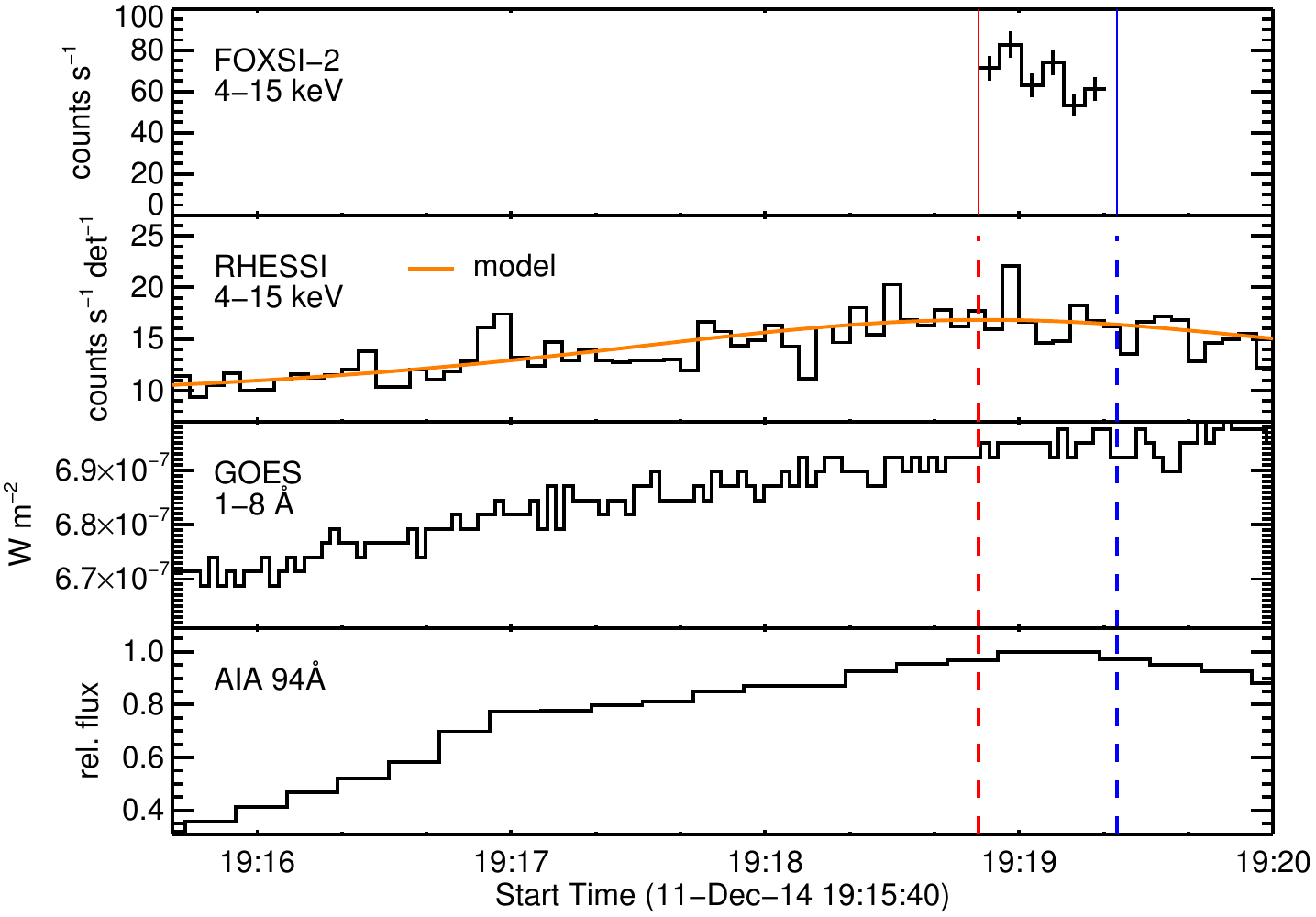}}
\subfigure{
\includegraphics[width=0.35\textwidth]{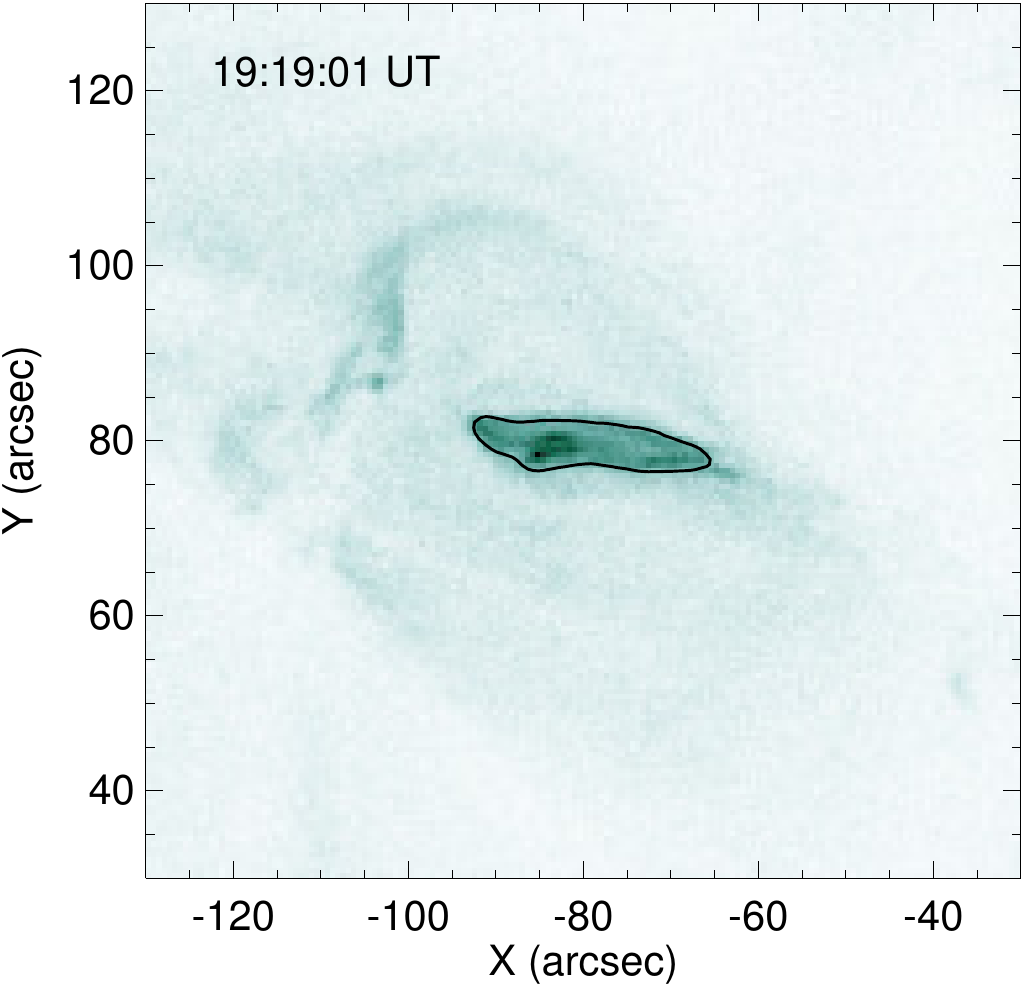}}
\caption{Light curves of microflare 2 from \textit{FOXSI-2}, \textit{RHESSI}, \textit{GOES}, and \textit{SDO}/AIA. This microflare was observed by \textit{FOXSI-2} (top panel) during the last target of the flight (Target J), denoted by red and blue vertical lines. \textit{FOXSI-2} data show the count rate within a circle (radius 100") centered on microflare 2, with corrections for vignetting effects. Data from only D6 are shown since attenuators were placed in front of all other detectors throughout the duration of this target.  The middle two panels show full Sun light curves from \textit{RHESSI} (4-15~keV), with the best fit light curve model (skewed Gaussian + linear background) overlaid, and \textit{GOES} (1.0-8.0~$\mathrm{\AA}$). The AIA light curve (94$\mathrm{\AA}$) is normalized to the maximum value and considers data from one dominant feature highlighted in the image on the right (intensity $>$ $30\%$). }
\label{fig:lc_mf2}
\end{figure}

\subsection{Timing Analysis}\label{sec:timing}

The \textit{FOXSI-2} light curves of microflares 1 and 2, shown in the top panel of Figures \ref{fig:lc_mf1} and \ref{fig:lc_mf2}, respectively, are created by selecting a circular region (radius 100") centered on the source of interest; the measured counts over the energy range 4-15~keV are then binned and corrected for vignetting. The instrument background is very low (${\sim}0.1$ counts s$^{-1}$ detector$^{-1}$ for the whole FOV), and so data are not background subtracted. For microflare 1, data from all five Si detectors are combined, while for microflare 2, data from only D6 are utlilized since attenuators were placed in front of all other detectors for a background measurement during Target J.  \\
\indent To better understand the overall evolution of these microflares, the \textit{FOXSI-2} light curves are compared to data from other X-ray instruments, \textit{RHESSI} (4-15~keV) and \textit{GOES} (1.0-8.0~$\mathrm{\AA}$), over the full Sun, and to data from \textit{SDO}/AIA. With the high angular resolution of AIA \citep[${\sim}1.5$";][]{lemen2012}, we can consider the evolution of individual features within the microflare regions. We use the AIA $94\mathrm{\AA}$ channel which captures flaring emission at ${\sim}6$~MK along with emission from lower temperature plasma at ${\sim}1$~MK. To isolate the Fe {\small XVIII} (higher-temperature) component of the $94\mathrm{\AA}$ channel, a linear combination of AIA channels as described in \cite{delzanna2013} can be utilized. We find no notable differences in the trends between the $94\mathrm{\AA}$ and Fe {\small XVIII} light curves for the studied microflares, indicating that the high-temperature component is dominant.    \\
\indent For microflare 1, we see that \textit{FOXSI-2} observed the declining phase of the flare during Targets A-E;  \textit{RHESSI} and \textit{GOES} data indicate that the flare began less than one minute prior to the start of \textit{FOXSI-2} observations (denoted by the red dotted line). In the \textit{RHESSI} data, there is a clear impulsive rise followed by a gradual decline. The impulsivity can be characterized quantitatively with an asymmetry index ($A_{sym}$), used in \cite{christe2008} and defined as, 

\begin{equation}
A_{sym} = \frac{t_{decay} - t_{rise}}{t_{decay} + t_{rise}},
\end{equation}

 \noindent where $t_{rise}$ and $t_{decay}$ are the rise and decay times of the flare, respectively. To estimate $t_{rise}$ and $t_{decay}$, we model the \textit{RHESSI} light curve data with a skewed Gaussian model plus linear background using the LMFIT package in Python \citep{newville2014}. The best fit model is overlaid on \textit{RHESSI} data in Figure \ref{fig:lc_mf1}. We then evaluate the skewed Gaussian component to find the time of the peak ($t_{peak}$), and we define the start and end times ($t_{start}$ and $t_{end}$) to be when the model component is at 5\% of the maximum count rate. Computing the rise and decay times as $t_{rise} = t_{peak} - t_{start}$ and $t_{decay} = t_{end}-t_{peak}$, we find the asymmetry index for microflare 1 to be $A_{sym}=0.65 \pm 0.06$. The positive value for $A_{sym}$ is indicative of an impulsive profile, confirming what we observe by eye. This quality is typical for HXR microflare emission; ${\sim}81\%$ of the \textit{RHESSI} microflares studied in \cite{christe2008} were found to be impulsive.  \\
\indent In the AIA $94\AA$ channel for microflare 1, a gradual rise in emission is observed, delayed compared to the \textit{FOXSI-2} and \textit{RHESSI} data due to its lower temperature coverage and the ionization timescales. Considering the AIA image in Figure \ref{fig:lc_mf1}, we see that multiple features are involved in microflare 1, which brighten and decay at different times throughout our observation. The evolution of these features will be explored further in Section \ref{sec:complexity}.  \\
\indent Microflare 2 shows distinct temporal characteristics compared to microflare 1, with context data from \textit{RHESSI}, \textit{GOES}, and AIA all showing a slow gradual rise in emission over the course of a few minutes. By studying the evolution of multiple regions on the Sun with AIA data, it is found that AR 12234 (microflare 2) is the only bright region showing a substantial rise in emission during the plotted time period, indicating that the rise in full Sun emission from \textit{GOES} and \textit{RHESSI} can be mostly attributed to this flare. Performing the same time series analysis with \textit{RHESSI} data as for microflare 1, we find that microflare 2 has an asymmetry index of $A_{sym}=-0.12 \pm 0.09$, indicating a more gradual time profile. With this non-impulsive profile, we note that it is more challenging to characterize the background for this flare. \\
\indent We additionally make a comparison between the \textit{RHESSI} and \textit{FOXSI-2} count rates as a consistency check.  To estimate the expected \textit{FOXSI-2} count rate from \textit{RHESSI} data, we first compute \textit{RHESSI} light curves for 1~keV intervals from 4-15 keV. Each curve is then adjusted for the instrument responses of \textit{RHESSI}\footnote{The \textit{RHESSI} instrument response includes a degradation factor determined through a comparison to \textit{GOES} flux available at https://hesperia.gsfc.nasa.gov/rhessi3/mission/operations/detector-efficiency/.} and \textit{FOXSI-2} before being combined into an integrated 4-15~keV light curve. After performing the modeling described previously on this adjusted curve, we evaluate the skewed Gaussian component to determine the expected \textit{FOXSI-2} count rate for each microflare. For microflare 1, we find an expected rate of $205\pm20$~counts~s$^{-1}$ at the beginning of Target A, which is consistent with the observed \textit{FOXSI-2} count rate of $217\pm12$~counts~s$^{-1}$. For microflare 2, the expected count rate at the beginning of Target J is $112\pm22$~counts~s$^{-1}$, which exceeds the observed \textit{FOXSI-2} rate of $71\pm6$~counts~s$^{-1}$. This discrepancy results from the challenge of characterizing the rising background observed in the \textit{RHESSI} data for this particular flare.

\begin{figure}
\centering
\includegraphics[width=0.46\textwidth]{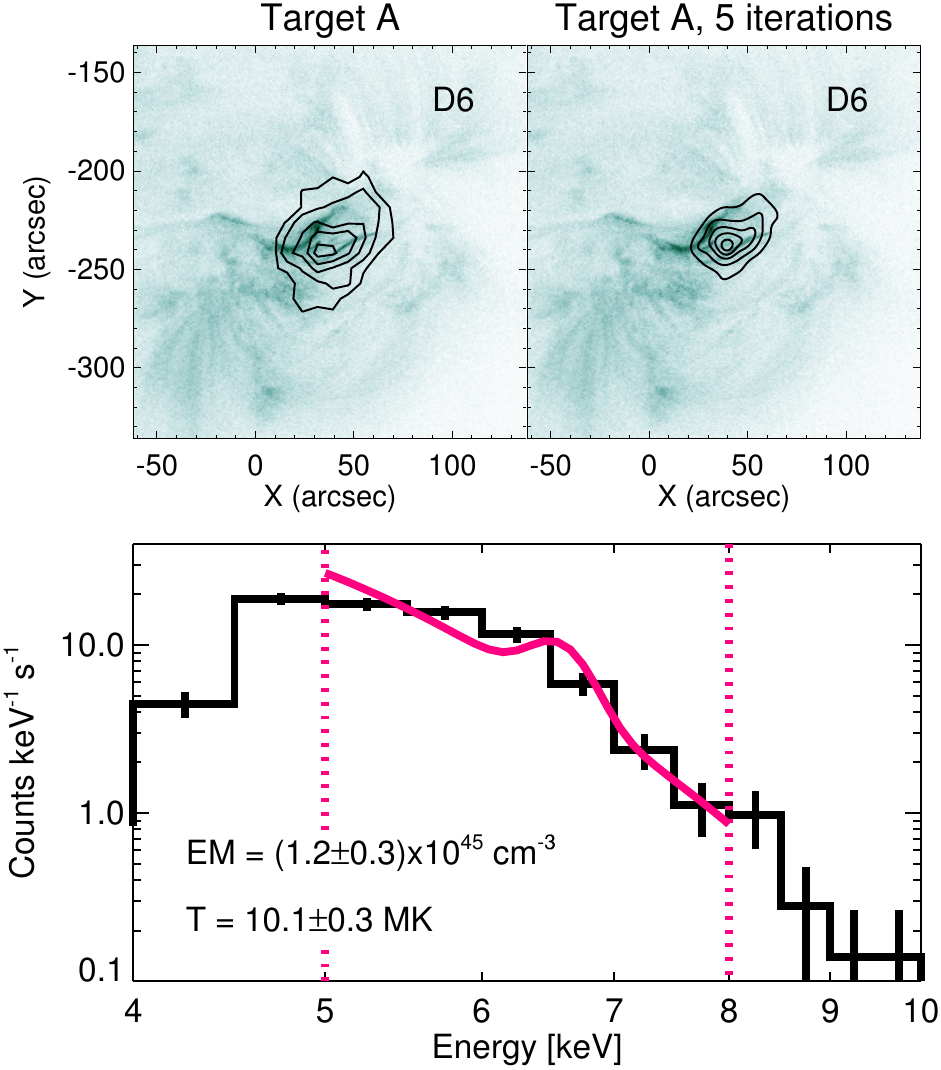}
\caption{\textit{FOXSI-2} images and spectrum for microflare 1 (Target A; duration $\sim 32$~s) using data from D6. Images show AIA $94\mathrm{\AA}$ data with a raw \textit{FOXSI-2} image (left) and a deconvolved \textit{FOXSI-2} image (right) overlaid (contours: 15\%, 30\%, 50\%, 70\%, 90\%). The \textit{FOXSI-2} images show only a portion of the FOV and include events in the energy range 4-15~keV. For the corresponding \textit{FOXSI-2} spectrum, an optically thin isothermal plasma model (magenta) is fit to the data (black) in the energy range 5-8~keV with bin size 0.5~keV. }
\label{fig:flare1}
\end{figure}

\begin{figure}
\centering
\includegraphics[width=0.46\textwidth]{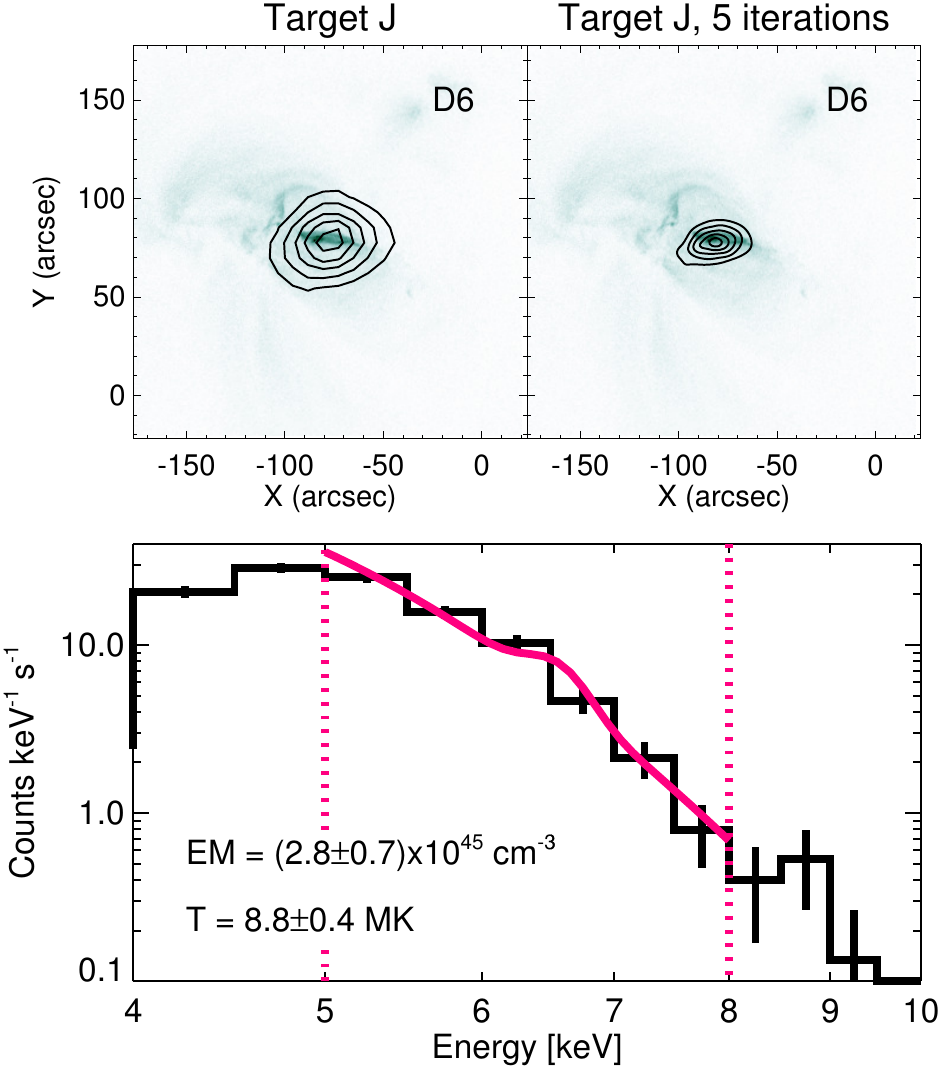}
\caption{\textit{FOXSI-2} images and spectrum for microflare 2 (Target J; duration $\sim 32$~s) using data from D6. Images show AIA $94\mathrm{\AA}$ data with a raw \textit{FOXSI-2} image (left) and a deconvolved \textit{FOXSI-2} image (right) overlaid (contours: 15\%, 30\%, 50\%, 70\%, 90\%). The \textit{FOXSI-2} images show only a portion of the FOV and include events in the energy range 4-15~keV. For the corresponding \textit{FOXSI-2} spectrum, an optically thin isothermal plasma model (magenta) is fit to the data (black) in the energy range 5-8~keV with bin size 0.5~keV. }
\label{fig:flare2}
\end{figure}

\subsection{Spectral Analysis}\label{sec:spec}

Spectral analysis was performed for each microflare using OSPEX\footnote{https://hesperia.gsfc.nasa.gov/ssw/packages/spex/doc/ospex\_explanation.htm} software, with the best fit model determined via chi-square minimization. A circular region with a radius of 100" centered on the source is selected for analysis. Given the narrow point spread function (PSF) of \textit{FOXSI}, we note that this is a conservative choice in order to include all photons affiliated with the microflare; at 100", the measured \textit{FOXSI} PSF shows that the relative flux is $\lesssim 10^{-4}$ compared to the on-axis source flux \citep{krucker2013}. The energy range is restricted to 5-8~keV due to uncertainty in the low-energy trigger efficiency below 5~keV and low statistics above 8~keV. \\
\indent Because of \textit{FOXSI-2}'s low instrument background, subtracting the background prior to spectral analysis for these microflares has no significant effect on the resulting parameters. Though singly-reflected photons \citep[``ghost rays";][]{peterson1997,milo2017} originating from bright regions on the eastern limb are likely present in the microflare observations, their contribution to the microflare spectra is expected to be small compared to the active region background. Spectral analysis on a background-subtracted spectrum was performed for microflare 2, using Target C as the background interval, as justified in \cite{ishikawa2017}. The resulting spectral parameters were consistent with those of the non-background-subtracted spectrum within uncertainty. The presented spectra in this paper are not background subtracted. \\
\indent The instrument response used for the analysis includes the optics effective area, absorption by thermal blanketing, and detector efficiency, which are all energy dependent; the energy resolution of the instrument, roughly constant across the \textit{FOXSI} energy range, is also incorporated (full width half maximum $\sim0.5~$keV for the Si detectors). Spectral analysis is performed separately for each telescope so that we can leverage the multiple measurements for investigating differences between telescopes. \\ 
\indent For each target on microflare 1 (Targets A-E) and for the only target during microflare 2 (Target J), an optically thin isothermal plasma model is fit to the data with the emission measure $EM$ and temperature $T$ as free parameters and the solar coronal abundances are fixed. The results shown in Figures \ref{fig:flare1} (microflare 1) and \ref{fig:flare2} (microflare 2) use data from a Si detector (D6) paired with a 10-shell optical module, which has a higher effective area, resulting in the best statistics out of all the detector-optic pairs. \\
\indent For microflare 1, Figure \ref{fig:flare1} is presented as a sample spectrum from the event; given that all the detectors were operating without the attenuator during microflare 1 (Targets A-E), we can utilize information from multiple detectors to assess the quality of our results. For each pointing, the parameters $T$ and $EM$ determined by spectral fitting of data from Si detectors\footnote{D4 is the only \textit{FOXSI-2} Si detector not included in the results for spectral analysis due to a currently incomplete understanding of the spectral shape of the response for this module.} D0, D1, D5, and D6 are combined as a weighted mean, presented in Table \ref{param}. These results show a decrease in emission measure over time, which is consistent with our understanding that \textit{FOXSI-2} was observing the declining phase of the flare. 

\begin{table}
\centering
\begin{tabular}{|c|c|c|}
\hline
\textit{Target} & \textit{T} (MK)         & \textit{EM} ($10^{44}$~cm$^{-3}$) \\ \hline
A    & $10.6 \pm 0.2$ & $7.6 \pm 1.0$                 \\
B     & $9.7 \pm 0.3$ & $7.7 \pm 1.5$                 \\
C    & $10.3 \pm 0.2$ & $5.3 \pm 0.7$                 \\ 
D     & $10.6 \pm 0.4$ & $5.3 \pm 1.3$                 \\
E   & $9.8 \pm 0.3$ & $2.9 \pm 0.6$    \\  \hline            
\end{tabular}
\caption{The weighted means of parameters from an optically thin isothermal plasma model fit to data from each of four Si detectors (D0, D1, D5, D6) during each target on microflare 1.}
\label{param}
\end{table}

\begin{figure}
\centering
\includegraphics[width=0.8\textwidth]{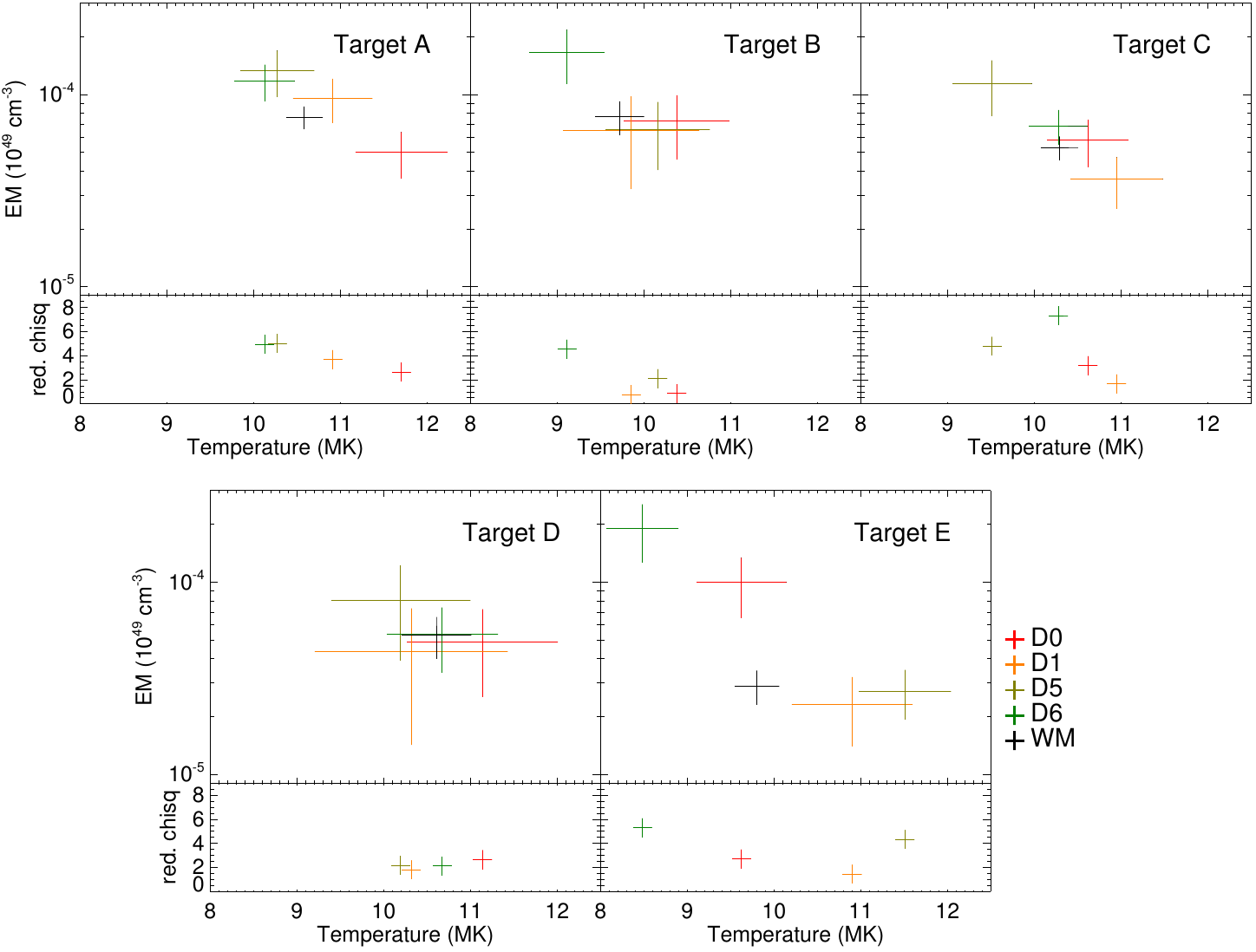}
\caption{Each of the five plots (Targets A-E) shows the emission measure vs. temperature from spectral fitting of data from each of four Si detectors during microflare 1 along with the weighted means (WM). The bottom panel of each plot shows the $\chi_{red}^2$ value for each fit. Several of these $\chi_{red}^2$ values are much larger than 1; this may be due to the fact we currently only include statistical error from the counts during spectral fitting, leaving out systematic error which we have yet to quantify. We note that the exposure time for Target D is relatively short ($\sim$10~s), resulting in lower statistics for these spectra. For Target E, the large off-axis position of the source may contribute to the large variation in derived spectral parameters.}
\label{fig:all_det}
\end{figure}

\indent For comparison, we plot the spectral parameters (\textit{EM} vs. $T$) for each detector and the weighted mean in Figure \ref{fig:all_det}. We note that, while the parameters are consistent between some detectors, there are instances where the error bars do not account for the spread in values. These discrepancies, along with the large $\chi_{red}^2$ values for some fits, may result from the fact that only statistical error from the measured counts are included in the spectral analysis, leaving out systematic error. By quantifying the observed variation in parameters compared to what is expected according to the parameter uncertainties, we can estimate what level of systematic error exists in the \textit{FOXSI-2} response. For these estimates, we consider the spectral parameters for Targets A-D; Target E shows a much larger variation in parameters than the other targets due to the large off-axis angle of the source. Because of the steepness of the off-axis vignetting curve in that part of the FOV, the response is sensitive to small shifts in off-axis angle which may be slightly different from telescope to telescope based on limits in alignment precision.

\indent The expected variations in $EM$ and $T$ between detectors are estimated by averaging the parameter uncertainties of the four considered detectors (representing the error due to statistics) while the observed variation is calculated as the standard deviation of the best fit parameters for all the detectors (representing the total error). For the plasma temperature, we find that the observed variation is roughly consistent with the expected variation of ${\sim}5\%$. However, for the emission measure, we observe ${\sim}50\%$ variation whereas the expected variation is only ${\sim}35\%$. This comparison indicates that, while the spectral shape of the instrument response is well-determined, there is systematic error in the relative normalization of the response between telescopes that we have not yet accounted for. \\
\indent To isolate the variation in emission measure, we fix the temperature to the weighted mean value, rerun the spectral analysis, and perform the calculation of expected and observed variation in $EM$ described above. In this case, the observed variation is ${\sim}18\%$ and the expected variation is ${\sim}9\%$. If we assume that the total variation corresponds to the random and systematic error combined in quadrature, we estimate that a systematic error of ${\sim}15\%$ is needed to account for the differences between telescopes; this is a reasonable amount of error to incur from the measurements of the various response elements \citep[e.g.,][]{boerner2012}. \\
\indent Even with the observed variation in parameters, data from all detectors for microflare 1 show evidence for high-temperature plasma ${\sim}10$~MK and relatively low emission measures below ${\sim}10^{45}$~cm$^{-3}$. This opens a novel parameter space for hard X-ray solar spectroscopic imagers, which will be discussed further in section \ref{sec:mic_compare}. 

\begin{figure}
\centering
\subfigure{
\includegraphics[width=0.95\textwidth]{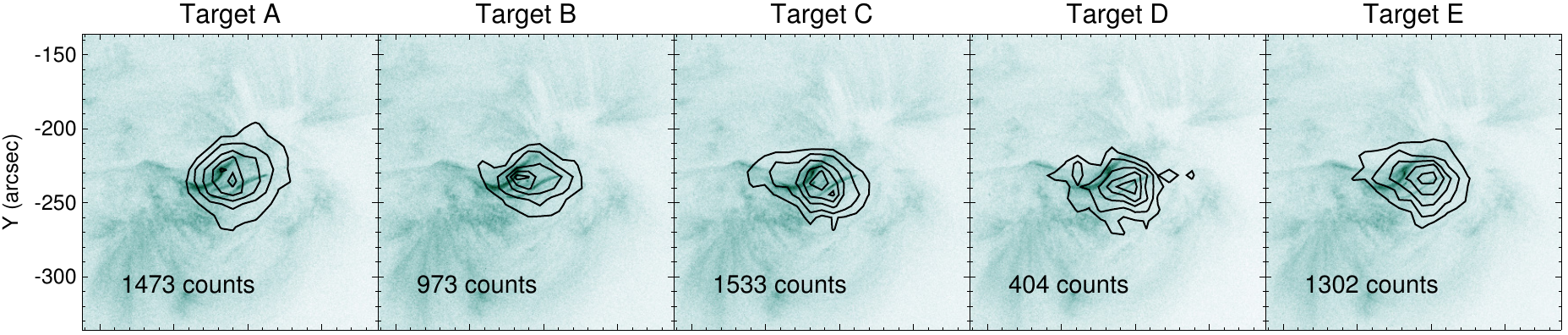}}\vspace{-1.5em}
\subfigure{
\includegraphics[width=0.95\textwidth]{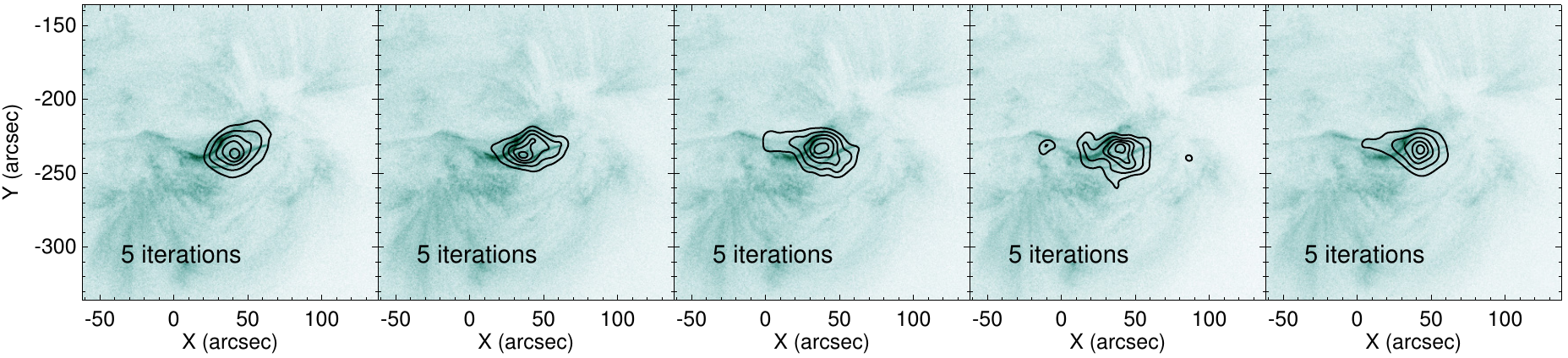}}
\caption{Images show microflare 1 during the first five targets (A-E) of \textit{FOXSI-2} over a portion of the FOV using data from all Si detectors (contours: 15\%, 30\%, 50\%, 70\%, 90\%) overlaid on AIA $94\mathrm{\AA}$ images. The top row shows the raw \textit{FOXSI-2} Si data, coregistered and added together for each target, along with the total number of counts (not scaled for exposure). The bottom row shows the results of a custom deconvolution method, described in detail in Appendix \ref{sec:deconv}, after 5 iterations.}
\label{fig:closeup}
\end{figure}

\indent From the derived spectral parameters, we can calculate what the \textit{GOES} series of spacecraft would expect to observe from such a flare and hence estimate the \textit{GOES} class. Using the weighted mean values of $EM$ and $T$ from Target A on microflare 1, as this interval is closest to the peak of the flare, we estimate a \textit{GOES} class of A0.1. For microflare 2, the \textit{GOES} class is estimated to be only slightly larger, at A0.3. These \textit{GOES} class estimates are of the same order as the estimates from Paper I using a multi-thermal DEM. \\
\indent Although use of an isothermal model can be useful for comparing our microflares to results from other X-ray instruments, we note that this type of model provides a limited picture of the events studied. For more comprehensive plasma characterization, computing the differential emission measure across a broad range of coronal temperatures is desired; this analysis of the \textit{FOXSI-2} microflares, along with the development of the \textit{FOXSI-2} temperature response, is detailed in Paper I. 

\begin{figure}
\centering
\includegraphics[width=0.95\textwidth]{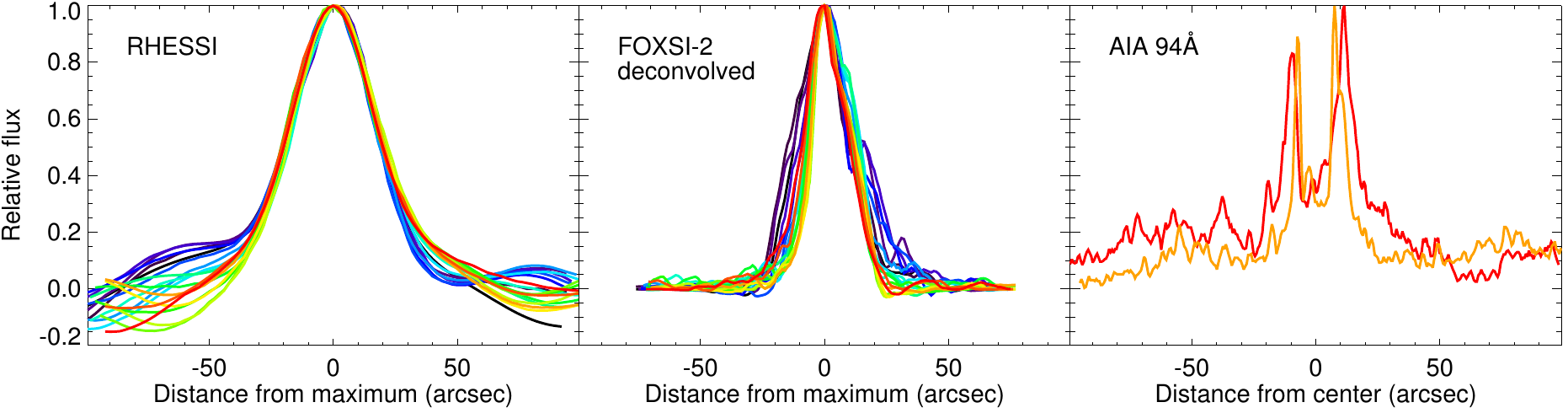}
\caption{Flux profiles of \textit{RHESSI} (left), deconvolved \textit{FOXSI-2} (middle), and AIA 94{\AA} (right) images of microflare 1 (Target A). For \textit{RHESSI} and \textit{FOXSI-2}, each curve represents the intensity measured along a line running through the center of the source at several angles ranging from 0 to 170 degrees. The AIA 94{\AA}  image shows flux profiles along the middle feature (red) and across the eastern and middle features (orange) as another indication  of the relevant size scales; for reference, the mentioned features are identified in Figure \ref{fig:lc_mf1}. We note that the \textit{RHESSI} CLEAN image (same as shown in Figure \ref{fig:rhessi}) does not utilize the high-resolution subcollimators due to the reasons to stated in Section \ref{sec:coalign}. However, we do not seek to compare the instrument resolutions with these flux profiles but rather to demonstrate the improvements gained in imaging dynamic range by using a direct imaging technique. With the \textit{RHESSI} image, the imaging noise extends up to ${\sim}15\%$ of the peak value at 1~arcmin, whereas the noise for the \textit{FOXSI-2} image is below $5\%$ of the peak at the same distance. }
\label{fig:rhessi_intensity}
\end{figure}

\subsection{Image Deconvolution}\label{sec:image_deconv}

The \textit{FOXSI} optics have a narrow, monotonically falling PSF, with a full width half maximum (FWHM) of ${\sim}5"$. We note that the detector resolution is coarse in comparison due to constraints on the focal length imposed by the sounding rocket payload size, with each strip crossing (or ``pixel") having a width of ${\sim}7.7"$ (${\sim}6.2"$) for the Si (CdTe) detectors. By characterizing the optics PSF through data collection at MSFC \citep{christe2016} and modeling, deconvolved images can be produced. \\ 
\indent We have developed a deconvolution method specifically for \textit{FOXSI} rocket data using a maximum likelihood procedure. In this method, a source map is convolved with the \textit{FOXSI} PSF and rotated to the detector plane for comparison to the measured data from each detector included in the analysis. After comparing the convolved source map to the measured data, adjustments are made to improve the source map over a set number of iterations. This method is described in greater detail in Appendix~\ref{sec:deconv}. \\
\indent Figure \ref{fig:closeup} shows the deconvolved images of microflare 1 during the first five targets (A-E) using this custom method. These images utilize data from all five Si detectors and show the source map after five iterations (chosen arbitrarily). When compared to the raw images, it is clear that the deconvolved images improve our ability to identify changes in morphology throughout the flare, such as the extension of emission towards the eastern feature in the AIA data starting during Target C. The evolution of microflare 1 will be explored in Section \ref{sec:complexity}. \\
\indent A comparison of the flux profiles of the \textit{RHESSI} and deconvolved \textit{FOXSI-2} images for microflare 1 (both with residuals added back in) highlights the improvement in imaging dynamic range that we gain by using a direct imaging technique. In Figure \ref{fig:rhessi_intensity}, each curve represents the intensity measured along a line running through the center of the source; this is measured for lines at a 10-degree increment ranging from 0 to 170 degrees for the \textit{RHESSI} and \textit{FOXSI-2} images. The AIA 94{\AA} image shows two flux profiles across the features of microflare 1 (see Figure \ref{fig:lc_mf1}) as context for the source extent. With the \textit{RHESSI} image, the imaging noise extends up to ${\sim}15\%$ of the peak value at 1~arcmin, whereas the noise for the \textit{FOXSI-2} image is below $5\%$ of the peak at the same distance.

\subsection{Imaging Spectroscopy}\label{sec:imspex}

\begin{figure}
\centering
\includegraphics[width=0.9\textwidth]{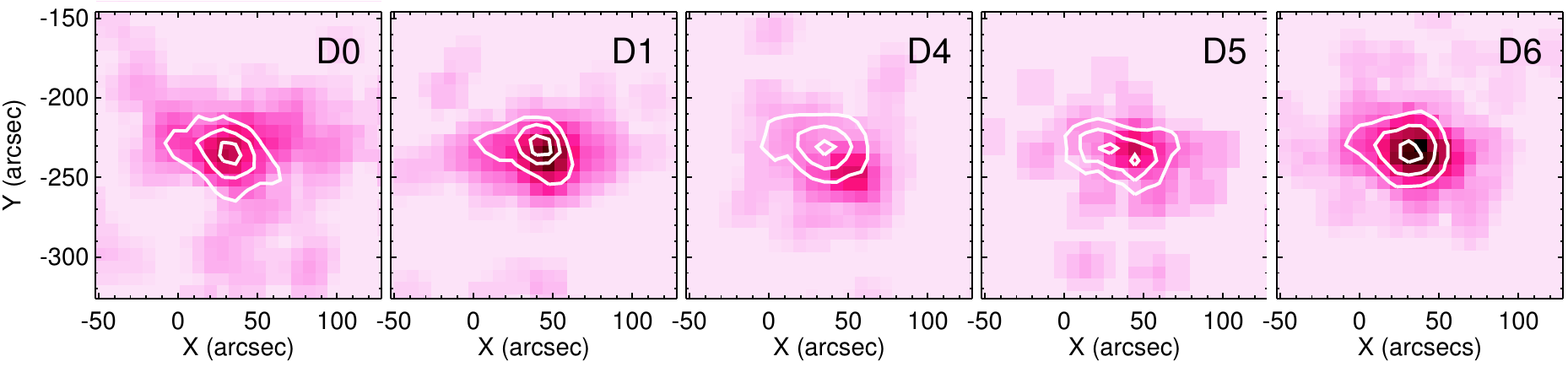}
\caption{Imaging spectroscopy for microflare 1 during Target C. The background image shows lower-energy data (4-5.5~keV) and contours show higher-energy data (6-15~keV) at 30\%, 60\%, and 90\% intensity. When calculating the image centroids for low- and high-energy emission, the high-energy emission is found to be east of the low-energy emission for each \textit{FOXSI-2} Si detector, with an average offset of ${\sim}7"$, roughly the size of one \textit{FOXSI} pixel. These results provide evidence for spatial complexity in a microflare of this size (discussed further in Section \ref{sec:complexity}).}
\label{fig:imspex}
\end{figure}

With the enhanced capabilities of \textit{FOXSI-2}, we are able to perform the first HXR direct imaging spectroscopy with finer angular resolution than \textit{NuSTAR} on a sub-A class flare. For this analysis, we select the target on microflare 1 where the source is closest to the center of the detector (Target C) since the effective area is highest towards the center due to vignetting effects. The counts are split into two energy bands: a lower energy band from 4-5.5~keV and a higher energy band from 6-15 keV, plotted in Figure \ref{fig:imspex} as the background image and contours, respectively. By calculating the centroids of both the low- and high-energy emission, it is found that the higher-energy emission is consistently offset to the east of the lower-energy emission for each Si detector, with an average offset of ${\sim}7$'' (roughly the width of one \textit{FOXSI} detector strip). This result suggests that there is higher temperature plasma in the eastern part of this flare.

\section{Discussion} \label{sec:discuss}

Spectral analyses of two sub-A class microflares observed by \textit{FOXSI-2} show evidence of flare-heated plasma at ${\sim}$10~MK and emission measures of ${\sim}10^{44}{-}10^{45}$~cm$^{-3}$, using an isothermal model. No clear evidence for a nonthermal component is observed for either flare; however, the possible parameter space for a hidden nonthermal component is explored in section \ref{sec:energetics}. Imaging spectroscopy shows a difference in plasma temperature over space within a sub-A class microflare, suggesting spatial complexity, which is discussed further in section \ref{sec:complexity} along with context data from \textit{SDO}/AIA.

\subsection{Comparing \textit{FOXSI-2} Microflares}\label{sec:mic_compare}

\begin{figure}
\centering
\subfigure{
\includegraphics[width=0.49\textwidth]{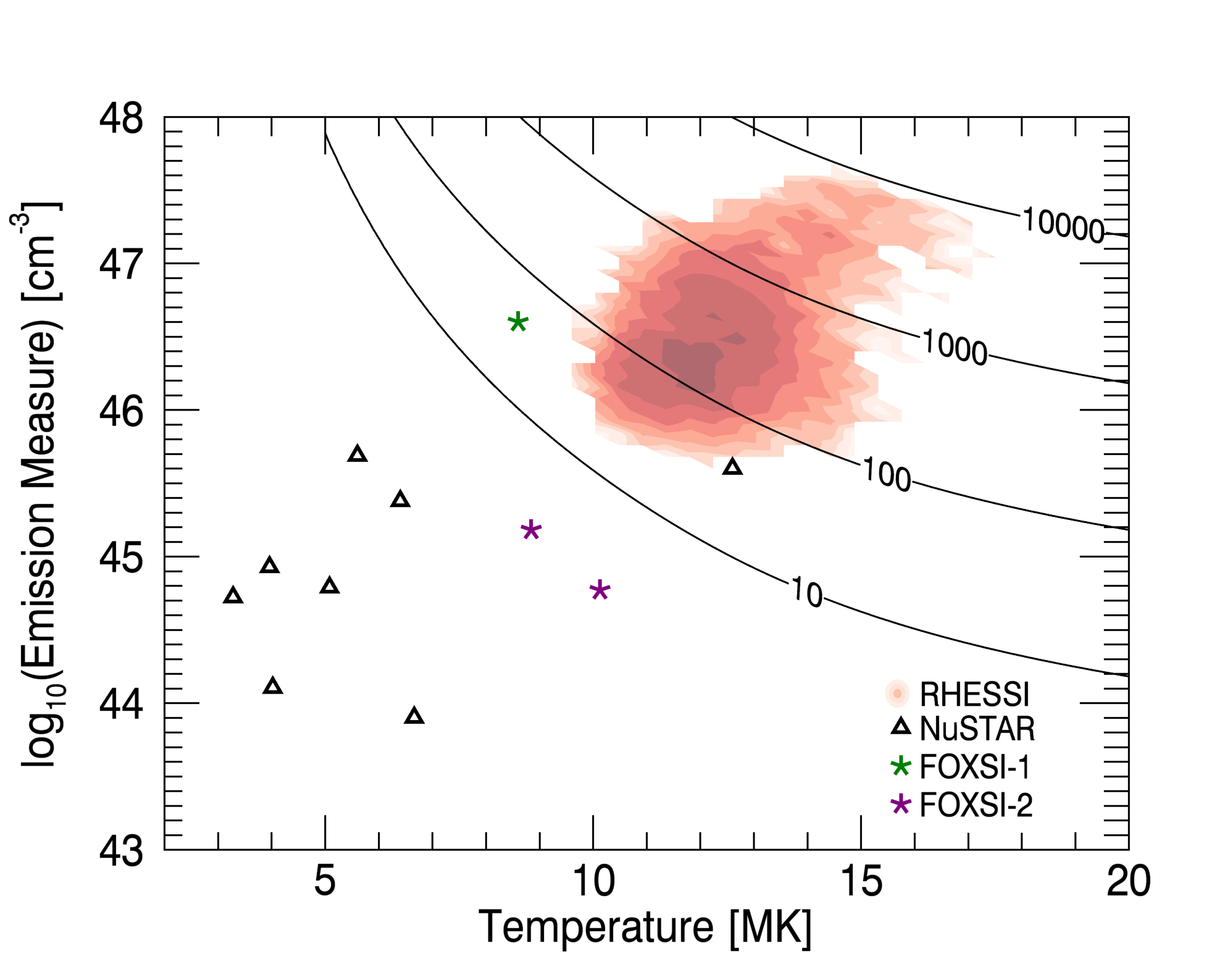}}\hspace{-2em}
\subfigure{
\includegraphics[width=0.49\textwidth]{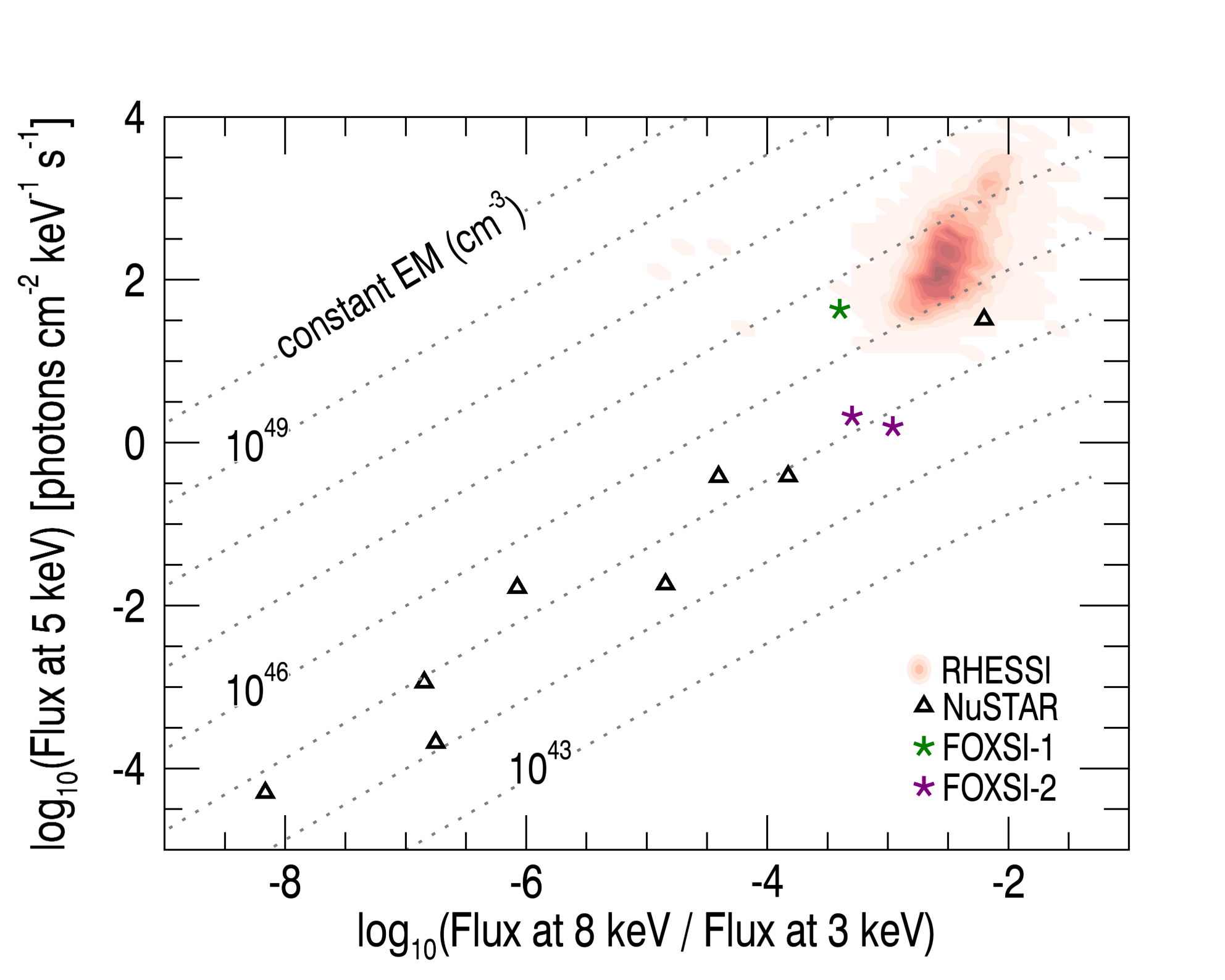}}
\caption{\textit{FOXSI} microflares are compared to solar microflares observed by other X-ray instruments. \textit{RHESSI} data (red) come from a comprehensive microflare study by \cite{hannah2008}. \textit{NuSTAR} data (black triangles) show flares from multiple studies, including microflares from \cite{glesener2017,glesener2020}, \cite{wright2017}, \cite{hannah2019}, and \cite{cooper2020} and three quiet Sun flares from \cite{kuhar2018}. The spectral parameters for the \textit{FOXSI-1} microflare (green star), shown here for the first time, were derived through analysis of \textit{FOXSI-1} and contemporaneous \textit{RHESSI} data. For \textit{FOXSI-2} (purple stars), we plot the weighted mean parameters for microflare 1 during Target A (see Table \ref{param}) while the results for microflare 2 use data from only D6 (see Figure \ref{fig:flare2}) due to the presence of attenuators in front of all the other detectors during Target J. The plot on the left, displaying $EM$ versus $T$, shows the best fit isothermal model for each flare. The plot on the right instead uses flux measurements as an analog to brightness and temperature, which allows for the inclusion of other models (e.g. double-thermal, nonthermal, etc.). These comparisons highlight how direct HXR spectroscopic imagers are opening up a novel parameter space for microflare studies. }
\label{fig:compare}
\end{figure}

With the spectral models derived in section \ref{sec:spec}, we can compare our \textit{FOXSI-2} microflares to microflares observed by other X-ray instruments on a plot of $EM$ vs. $T$ (see Figure \ref{fig:compare})  for isothermal models. We additionally plot the photon flux at 5 keV (photons cm$^{-2}$ keV$^{-1}$ s$^{-1}$) against the ratio of the flux at 8 keV to the flux at 3 keV; this representation serves as an analogue to microflare brightness versus temperature while allowing for the consideration of other models that may provide a better fit the data, including double-thermal and nonthermal models. In these comparison plots, we note that the \textit{FOXSI-2} microflares are roughly an order of magnitude fainter than the faintest microflares observed by \textit{RHESSI}. Overall, these plots highlight how direct HXR spectroscopic imagers are opening up a novel parameter space for high-energy solar microflare studies. \\ 
\indent We further note that \textit{FOXSI-2}'s sensitivity to high temperature plasma (${\sim}10$~MK) complements \textit{NuSTAR}'s sensitivity to lower temperature plasma. Though \textit{NuSTAR} has a larger effective area than \textit{FOXSI-2} ($800$~cm$^2$ versus $80$~cm$^2$ at ${\sim}10$~keV), \textit{FOXSI-2} is more sensitive to faint emission at higher energies due to its increased detector throughput. At the peak of a microflare, \textit{NuSTAR}'s livetime is typically reduced to ${\sim}1\%$ \citep[e.g.,][]{glesener2017}, while \textit{FOXSI-2}'s livetime remains at ${\sim}50\%$ for similar events.  As a result, \textit{FOXSI-2} achieves a sensitivity that is $\frac{80 cm^2}{800 cm^2} \cdot \frac{50\%}{1\%} \approx 5$ times greater than that of \textit{NuSTAR} above 10~keV for typical microflare observations.

\subsection{Flare Energetics}\label{sec:energetics}

One question to consider when studying small-scale solar flares is: are small flares similar in structure to large flares, just scaled down in size? This can be answered, in part, by checking if microflare energetics follow the standard flare model; if these flares are consistent with the standard model, we would expect the amount of energy in nonthermal electrons to be large enough to account for the thermal energy in the flare. We can investigate this by making estimates of and comparing the thermal and nonthermal energies. Though no clear evidence for a nonthermal component was observed, we can still estimate the possible nonthermal energy available by seeing how large a nonthermal component could exist undetected below the thermal model. \\
\indent The thermal energy $E_{therm}$ is estimated using the spectral parameters $T$ and $EM$ from our isothermal fits, such that 

\begin{equation}
E_{therm}=3kT \sqrt{EM \cdot V},
\end{equation}

\noindent where $V$ is the volume of the emission region. We utilize estimates of $V$ from Paper I, which were computed by first estimating the area of emission $A$ using AIA Fe {\small XVIII} maps and then setting $V{\sim}A^{3/2}$. From this computation, we obtain estimates of the thermal energy: $E_{therm} \sim (1.9~\pm~0.2)~\times~10^{28}$~erg for microflare 1 (Target A) and $E_{therm} \sim (1.3 \pm 0.2) \times 10^{28}$~erg for microflare 2. We note that the thermal energy estimates made using an isothermal model are consistently lower than those determined in Paper I using a multi-thermal DEM. \\
\indent To estimate the energy in nonthermal electrons, we add a fixed cold thick-target component (\texttt{thick2}) to the spectral model based on an electron spectrum $\Phi_e \sim \epsilon^{-\delta}$ with index $\delta$, low energy cutoff $E_c$, and integrated electron flux $R_e$ (electrons~s$^{-1}$) \citep{brown1971}. We set the nonthermal energy $E_{NT}$ to the thermal energy estimates from Paper I, $5.1 \times 10^{28}$~erg for microflare 1 and $1.6 \times 10^{28}$~erg for microflare 2. Values of $\delta$ ranging 3-15 ($\Delta\delta=0.5$) and $E_c$ ranging 3-10~keV ($\Delta E_c=0.5$~keV) are tested; each combination of $\delta$ and $E_c$ with the set value for $E_{NT}$ then constrains the value of $R_e$ such that 

\begin{equation}
R_{e}=\frac{\delta-2}{\delta-1} \frac{E_{NT}}{E_C \Delta t},
\end{equation}

\noindent where $\Delta t$ is the length of the observation. We note that this method provides only a limited subset of the possible parameter space, as we have restricted the study to electron spectrum parameters resulting in a nonthermal energy equal to the thermal energy. \\
\indent For microflare 1, we find that including a nonthermal component with $E_C\sim3~$keV and $\delta\sim7{-}8$ results in a small improvement in the fit compared to an isothermal model alone. In these cases, we find that there is no significant change in the thermal parameters, i.e., $T$ and $EM$ are consistent with that of the isothermal model within uncertainty. These electron spectrum parameters thus allow for a dominant thermal component while also providing enough energy in nonthermal electrons to account for the thermal energy, consistent with the cold thick-target model. \\
\indent The softer spectral indices found in this investigation are consistent with the trend we would expect for small-scale events based on scaling laws \citep[e.g.,][]{battaglia2005,isola2007}. Results from other microflare studies in the hard X-ray regime support these trends, including the population study in \cite{christe2008}, which finds that the electron spectra for microflares observed by \textit{RHESSI} have an average spectral index of $\delta=8.4$ with a standard deviation of 2.7. Additionally, a recent paper by \cite{glesener2020} finds evidence for a nonthermal component for a class A5.7 microflare observed by \textit{NuSTAR}, with the best fit model including an accelerated electron distribution characterized by $\delta=6.3 \pm 0.7$ and $E_C\lesssim6.5$~keV. The combination of a softer spectral index and lower low-energy cutoff compared to larger flares is consistent with the \textit{NuSTAR} microflare studies of \cite{wright2017} and \cite{cooper2020}, which consider upper limits to the nonthermal emission, and is consistent with what we find for our best fit model.\\
\indent With an accelerated electron spectrum extending down to lower energies, the warm thick-target model may be more appropriate. In a warm thick-target scenario, accelerated electrons with energies below a few times the average energy in the thermal population ($E_e = \frac{3}{2}kT$) will thermalize in the corona prior to reaching the flare footpoints and contribute to the observed thermal spectrum \citep{kontar2019}. With the best fit spectrum for microflare 1 ($\delta=7$, $E_C=3$~keV), the average energy of the nonthermal electrons is ${\sim}3.5$~keV, which is only a few times larger than the average energy of the thermal electrons (${\sim}1.3$~keV). \\
\indent To test this model, we utilize \texttt{thick\_warm} in OSPEX, leaving the parameters $R_e$, $\delta$, and $E_C$ free, while fixing the thermal parameters based on the isothermal model and estimates of the emission volume. Through this analysis, we find that the warm thick-target model provides a worse fit to the data, and the free parameters are poorly constrained. In particular, the fit fails to constrain $E_C$, because it is below the energy range where \textit{FOXSI-2} is sensitive. We attribute this poor fit to the multiple loops producing the observed emission for microflare 1, as \texttt{thick\_warm} is designed for a single flare loop. \\
\indent The nonthermal analysis described above is also applied for microflare 2. With a cold thick-target model, none of the tested combinations of $\delta$ and $E_C$ provide an improved fit compared to the isothermal model alone, indicating that this observation is less consistent with the standard model we observe for larger flares. The warm thick-target model also provides a worse fit to the data than the isothermal model, with the parameters again poorly constrained. Revisiting the light curves for microflare 2 (see Figure \ref{fig:lc_mf2}), we note that the peak is preceded by a gradual rise in emission over multiple minutes, compared to the sharp impulsive rise over ${\sim}20$~s observed by \textit{RHESSI} for microflare 1. In the study of \textit{RHESSI} microflares by \cite{christe2008}, though most microflares were found to be impulsive, ${\sim}18\%$ had a more gradual rise (rise time $>$ decay time), similar to microflare 2. Both the difference in time evolution and lack of evidence for a nonthermal component suggest that this flare develops in a manner that is distinct from the standard flare model.  

\subsection{Flare Complexity}\label{sec:complexity}
\begin{figure}
\centering
\includegraphics[width=0.56\textwidth]{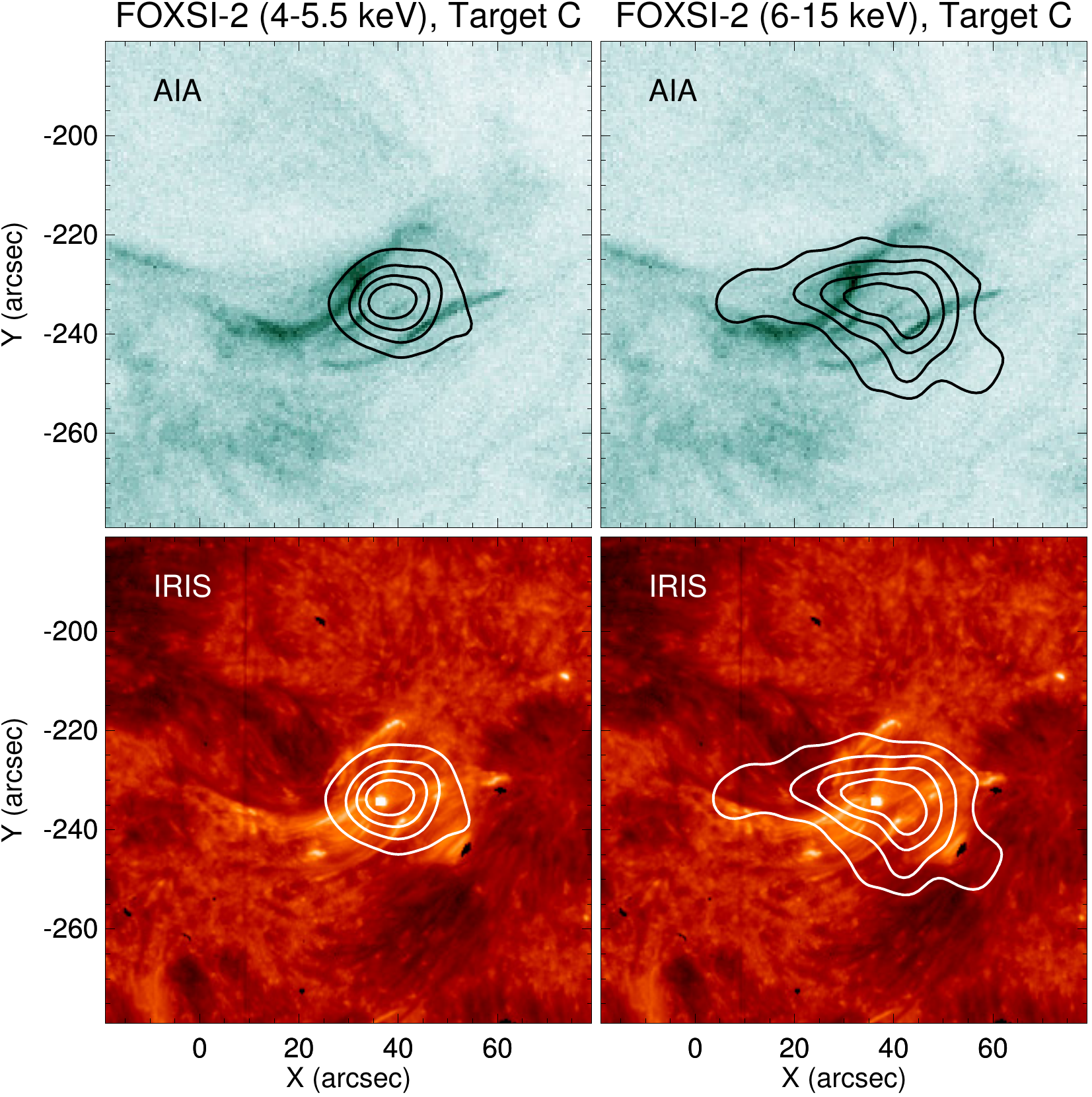}
\caption{Deconvolved \textit{FOXSI-2} images of microflare 1 (Target C) in two energy bands, 4-5.5~keV (left column) and 6-15~keV (right column), overlaid on contemporaneous AIA $94\mathrm{\AA}$ (top row) and \textit{IRIS} $1330\mathrm{\AA}$ (bottom row) images. There is a higher fraction of high- to low-energy emission extending out towards the eastern feature observed in AIA, indicating hotter plasma. This region of hotter plasma also overlaps with the \textit{IRIS} slit (at $x{\sim}10$"). }
\label{fig:energy_bands}
\end{figure}

In addition to studying the energetics, we are also interested in investigating the spatial and temporal complexity of solar microflares; at what point, if any, does a small-scale flare lose the complexity that we see in large solar flares and become a single-energy-release event, more similar to what we expect for nanoflares? From the imaging spectroscopy described in section \ref{sec:imspex}, there is evidence of plasma heated to different temperatures at different spatial locations throughout the flare. \\
\indent We further investigate flare dynamics by comparing our \textit{FOXSI-2} observations with contemporaneous \textit{SDO}/AIA data. Looking at the AIA light curves for microflare 1 in Figure \ref{fig:lc_mf1}, we see that the western feature brightens first and then fades as the middle feature brightens, followed by the brightening of a small eastern feature, showing both temporal and spatial complexity. We suggest that the observed difference in plasma temperatures from the \textit{FOXSI-2} data reflects the heating of the eastern and middle features while the western feature is cooling. \\
\indent We also observe these dynamics at play within the deconvolved \textit{FOXSI-2} images for microflare 1. In the series of deconvolved images in Figure \ref{fig:closeup}, we can clearly identify the extension of emission out toward the east starting to appear during Target C, just as AIA $94\mathrm{\AA}$ emission from the eastern feature is beginning to rise (Figure \ref{fig:lc_mf1}). To further explore the results from the imaging spectroscopy of Target C, we produced deconvolved images of the two studied energy-bands (4-5.5~keV and 6-15~keV). These images, presented in Figure \ref{fig:energy_bands}, show that a higher fraction of high-energy emission can be found in the region extending out to the small eastern feature in AIA $94\mathrm{\AA}$. This higher energy emission additionally overlaps with the slit in contemporaneous \textit{Interface Region Imaging Spectrograph} (\textit{IRIS}) data. A future paper will use \textit{FOXSI-2} data, \textit{IRIS} spectral data, and modeling to study the impact of an accelerated electron beam on the lower atmosphere.\\    
\indent The case for thermal complexity is additionally supported by the analysis of Paper I, where simultaneous brightenings were observed across a broad energy range. Thus the combination of \textit{FOXSI-2} data with contemporaneous data from instruments such as \textit{SDO}/AIA and \textit{Hinode}/XRT provides compelling evidence for complex energy release for this sub-A class flare, indicating that it is not a nanoflare as defined by \cite{parker1988}. Based on our analysis, this flare requires a minimum number of three energy releases. 

\newpage

\section{Summary} \label{sec:summary}

By utilizing a direct imaging technique, the \textit{FOXSI} sounding rocket experiment provides improved sensitivity and imaging dynamic range for small-scale solar events in the hard X-ray regime. During the \textit{FOXSI-2} flight, two microflares were observed; beyond simply detecting these sub-A class flares, we are able to perform detailed spectral and imaging analysis with \textit{FOXSI-2} data. This analysis reveals the presence of high temperature plasma (${\sim}10$~MK) and highlights \textit{FOXSI-2}'s improved sensitivity with emission measures that are an order of magnitude smaller than what was observed from the faintest \textit{RHESSI} flares. \\
\indent Multiple results from our analysis indicate that we can be confident in the characterization of the \textit{FOXSI-2} instrument response. First, we find that the measured \textit{FOXSI-2} count rate is consistent with that of \textit{RHESSI}, after accounting for the response of each instrument. Additionally, by leveraging the measurements from multiple \textit{FOXSI-2} telescopes, we are able to assess the level of systematic error in our experiment. A comparison of the derived spectral parameters ($T$, $EM$) between telescopes indicates that the spectral shape of the \textit{FOXSI-2} response is well-determined and that there is only ${\sim}15\%$ systematic error in the relative normalization of the response between detectors. \\     
\indent Through studies of the energetics of the \textit{FOXSI-2} microflares we find the thermal energies to be ${\sim}10^{28}$~erg, which is around the energy at which \textit{RHESSI}'s sensitivity starts to limit the fraction of observed events. Additionally, exploration of the parameter space for an electron spectrum that could provide enough nonthermal energy to account for the thermal energy in microflare 1 allows for spectra with $E_c\sim3~$keV and $\delta\sim7{-}8$. With these parameters, it is plausible that this microflare abides by the picture of energy transfer described in the standard model for solar flares. Furthermore, imaging spectroscopy of \textit{FOXSI-2} data and contemporaneous AIA data for microflare 1 provide evidence for spatial and temporal complexity, supporting the idea that this microflare more closely resembles the structure and dynamics of a large flare rather than the single energy release of a nanoflare. In the future, more high-sensitivity observations from solar hard X-ray instruments like \textit{FOXSI} can help us to better understand the characteristics of microflares and their contribution to coronal heating.       \\

\acknowledgments
The \textit{FOXSI} sounding rocket experiment is supported through NASA LCAS grants NNX08AH42G, NNX11AB75G, and NNX16AL60G. This work is additionally supported by an NSF Faculty Development Grant (AGS-1429512), an NSF CAREER award (NSF-AGS-1752268), the SolFER DRIVE center (80NSSC20K0627), and by NASA Headquarters under the NASA Earth and Space Science Fellowship Program (80NSSC17K0430). The authors would like to acknowledge the contributions of each member of the \textit{FOXSI} experiment team, particularly our team members at JAXA/ISAS for the provision of Si and CdTe detectors and at NASA/MSFC for the fabrication of the focusing optics. The authors also wish to acknowledge and thank Richard Schwartz for his contributions to the development of the \textit{FOXSI} image deconvolution method described in the appendix.

\appendix

\section{Image Deconvolution}\label{sec:deconv}

\subsection{Custom \textit{FOXSI} Deconvolution Method}

\begin{figure}[b]
\centering
\includegraphics[width=0.35\textwidth]{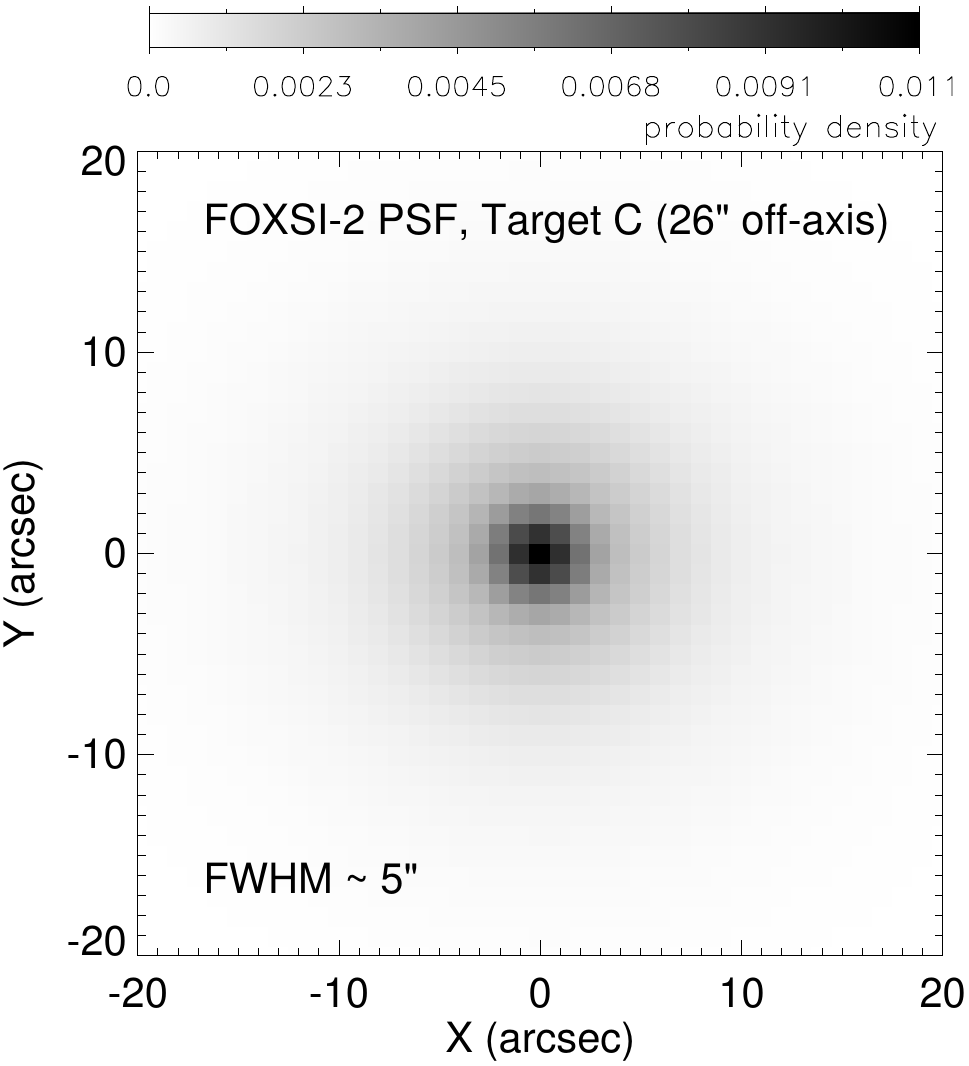}
\caption{\textit{FOXSI-2} PSF map used for the deconvolution of microflare 1 images during Target C (source centroid ${\sim}26"$ off-axis). The PSF is incorporated in the instrument response $S$ during Step \ref{deconv_step1} of the deconvolution procedure.}
\label{fig:psf}
\end{figure}

The \textit{FOXSI} image deconvolution method utilizes a maximum likelihood procedure \citep{richardson1972,lucy1974,benvenuto2013}, which aims to derive the source map $W$ ($j$ pixels) from a set of observations $H$ ($i$ pixels) by accounting for the instrument response $S$. This is achieved by iterating over the following equation:

\begin{equation}
W_j^{t+1} = W_j^{t} \sum_{i} \frac{H_i}{C_i} S_{ji},
\label{eq:iter}
\end{equation}

\noindent where $C$ is the reconvolved source map, such that $C_i = \sum_{k} S_{ki} W_k^t$. For each iteration $t$, the ratio $H_i/C_i$ indicates where the reconvolved source map overestimates or underestimates the observations, allowing for a correction to be made in the next iteration of the source map, $W_j^{t+1}$. We note that this procedure allows for the source map and observations to have different bases; this capability is important for \textit{FOXSI} data as it allows for the source map to have a much finer binning than \textit{FOXSI's} coarse strip crossings (${\sim}7.7"$ for Si). Furthermore, $H$ can include observations from multiple detectors. \\
\indent The instrument response $S$ incorporates the optics PSF, the detector rotation(s), and the difference in bin size between the source map and the raw \textit{FOXSI} images. Currently, the deconvolution method uses an invariant PSF, as the PSF does not change much over the relatively small FOV considered for the sources of interest ($\sim 2' \times 2'$ for this study). The PSF is modeled as three two-dimensional Gaussians based on measurements made at the MSFC Stray Light Facility for on-axis and multiple off-axis source angles \citep{christe2016}. The PSF is determined for any given position in the FOV through interpolation (see Figure \ref{fig:psf} for an example PSF). \\
\indent The deconvolution method follows the following steps:

\begin{enumerate}
	\item{Compute the instrument response $S_{ji}$ for a given PSF map and set of detectors. In this computationally intensive procedure, the predicted counts in each bin $i$ is calculated for a source $W_j$. The source at position $j$ is convolved with the optics PSF, rotated to the detector plane (for each given detector roll angle), and rebinned to the detector strip crossing size.}\label{deconv_step1}
	\item{Retrieve observations $H_i$ for the set of detectors specified in Step \ref{deconv_step1} over a given energy range (4-15~keV in this study). }
	\item{Perform the iterative procedure defined in Equation \ref{eq:iter}, starting with a gray-scale array for the initial source map $W^0$. An arbitrary stopping point of 5 iterations has been chosen for this study based on improvements in the source image.}
\end{enumerate}

\begin{figure}
\centering
\includegraphics[width=0.9\textwidth]{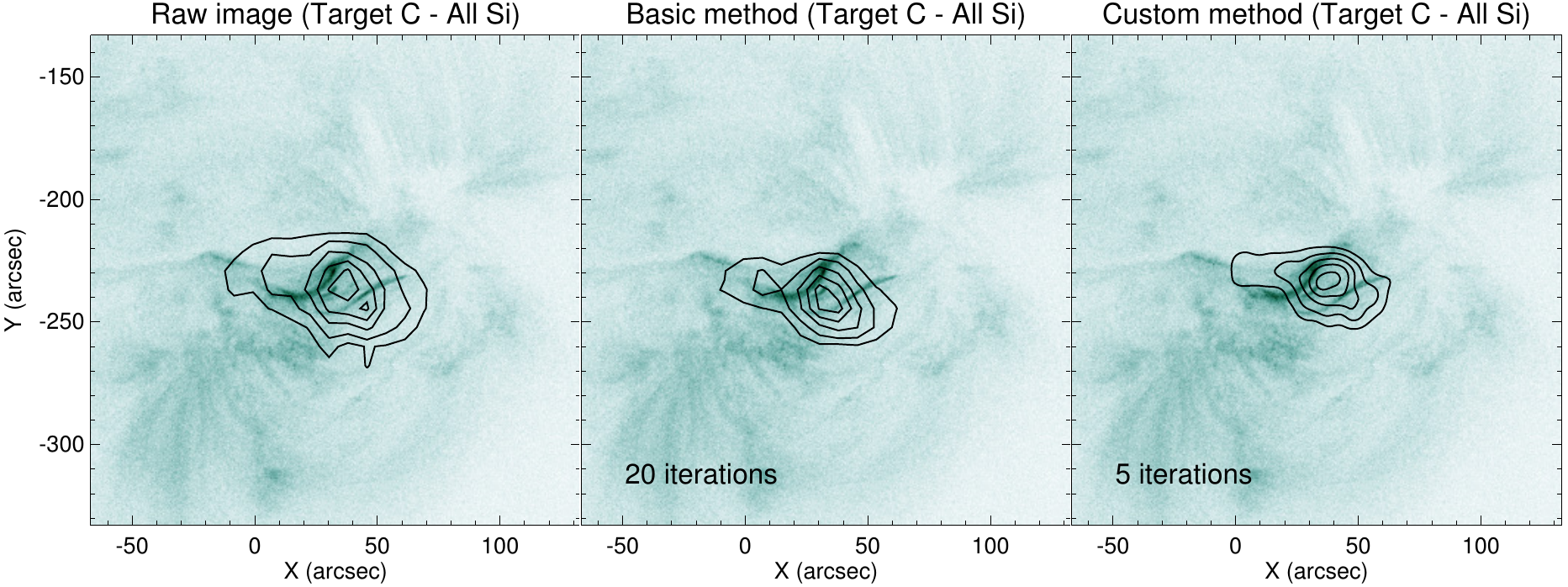}
\caption{Comparison of deconvolution methods for \textit{FOXSI-2} images. (\textit{Left}) Raw \textit{FOXSI-2} image of microflare 1 during Target C, with data from all Si detectors coregistered and combined. (\textit{Middle}) Deconvolved \textit{FOXSI-2} image after 20 iterations using a standard deconvolution method (max\_likelihood.pro). (\textit{Right}) Deconvolved \textit{FOXSI-2} image after 5 iterations using a custom deconvolution method. Though both methods reveal similar features overall, the custom method captures finer details of the flare structure. The contours represent intensities of 15\%, 30\%, 50\%, 70\%, and 90\%.} 
\label{fig:compare_methods}
\end{figure}

\begin{figure}
\centering
\subfigure{
\includegraphics[width=0.87\textwidth]{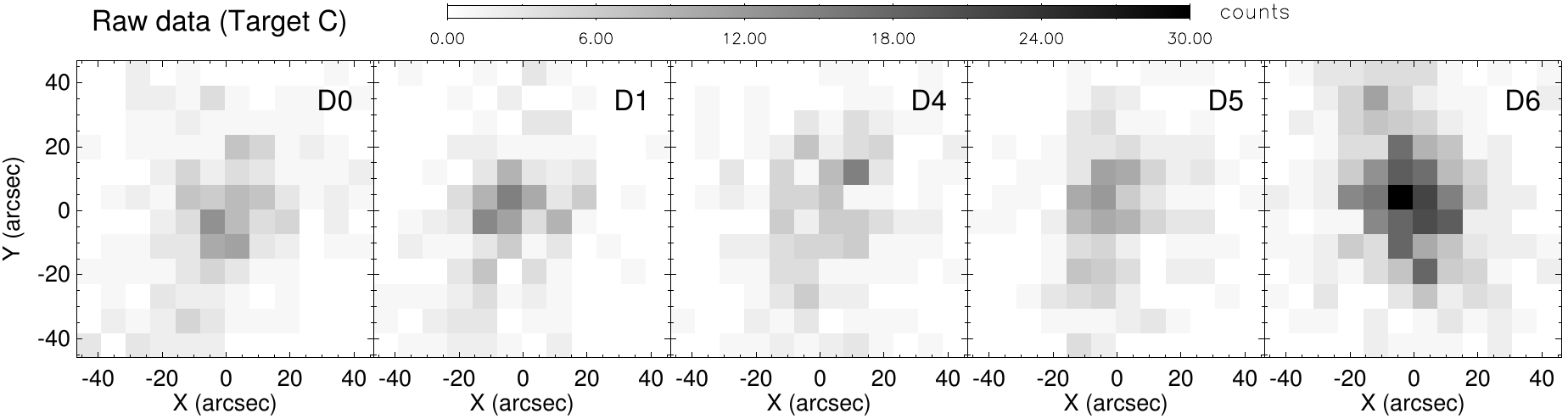}}
\subfigure{
\includegraphics[width=0.87\textwidth]{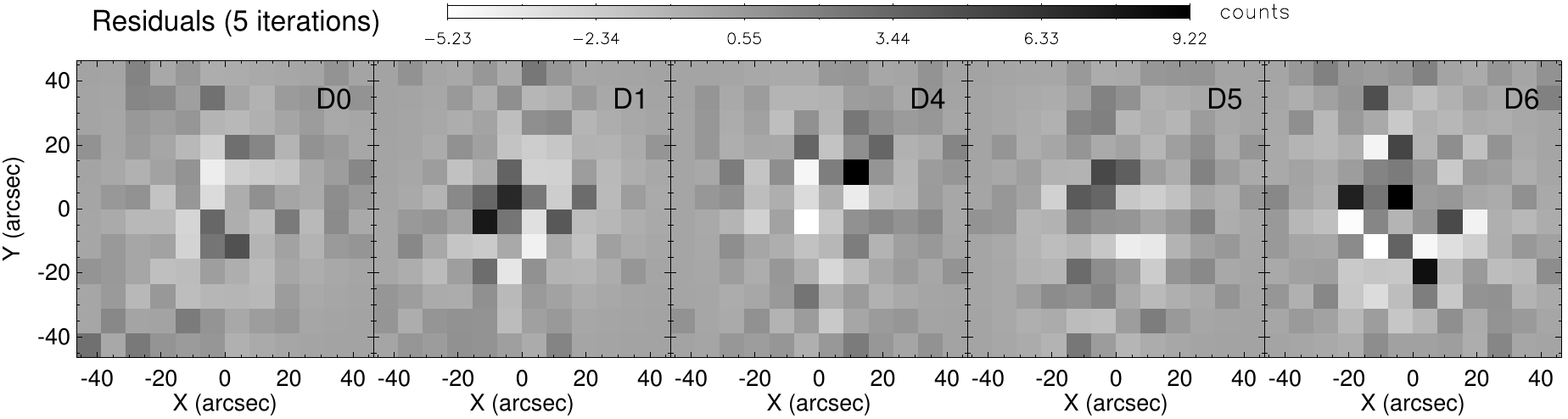}}
\subfigure{
\includegraphics[width=0.59\textwidth]{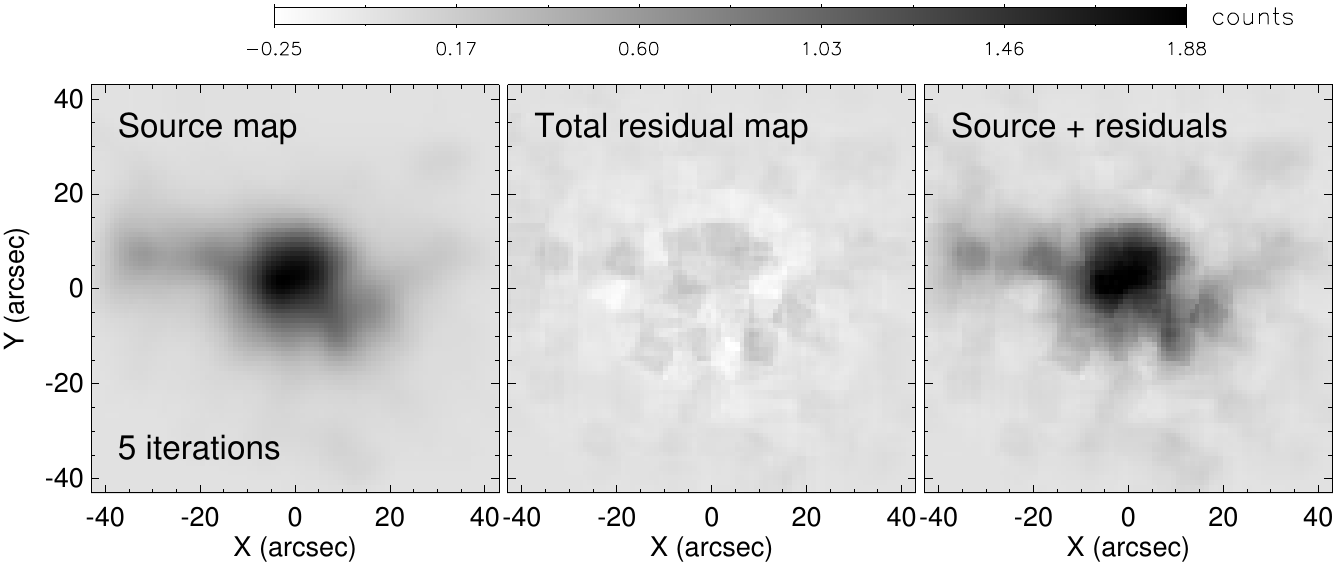}}
\caption{(\textit{Top}) Raw \textit{FOXSI-2} images in the detector plane for the five Si detectors during microflare 1 (Target C). Each detector is rotated at a different angle within the focal plane and thus has a different orientation relative to solar north. D6 shows a higher count total compared to the other detectors due its paired optic (10-shell), which has a higher effective area. (\textit{Middle}) Residual maps for each Si detector in the detector plane during Target C after 5 iterations. The lack of a distinct pattern in the residuals indicates that the source map reasonably represents the observations, and no major source region is missing. (\textit{Bottom}) The source map (left), total residual map (middle), and combined (source + residuals) map (right) for Target C after 5 iterations, all aligned with solar north. In the total residual map and combined (source + residuals) map, we observe artifacts of the coarse \textit{FOXSI-2} strip crossings (7.7"), which are large compared to the source pixels (1"). }
\label{fig:resid}
\end{figure}

The results of the deconvolution for microflare 1 using this custom method are compared to that of a standard method (max\_likelihood.pro) and to a raw \textit{FOXSI-2} image in Figure \ref{fig:compare_methods}. The raw image is created by coregistering and combining images from each Si detector during Target C. This image serves as the observation map for the standard method, which requires that the source map and observations have the same basis. Consequently, the resulting source map is constrained to the coarse resolution of the \textit{FOXSI-2} Si detectors. Additionally, we introduce error to the observations by rotating, rebinning, and combining detector images. The custom method avoids this problem by leaving the observations in each detector plane (see top row of images in Figure \ref{fig:resid}) and allows for a finer resolution source map. With these improvements, we can better identify and trace changes in the flare structure. 

\subsection{Residual Maps}\label{sec:resid}

In the deconvolution process, a residual map can be calculated by subtracting the reconvolved source map from the observed map for a given iteration. These residuals are useful for determining how well the source map accounts for observations and can be added back into the source map as a way to probe the instrument's imaging dynamic range. \\
\indent To produce a residual map for a given iteration of the \textit{FOXSI} image deconvolution method, the following steps are implemented:

\begin{enumerate}
	\item{Calculate the residuals for each individual detector included in the analysis. In the detector plane, the counts from the reconvolved source map are subtracted from the observed counts for each strip crossing.}
	\item{Rebin and rotate the residual map for each detector to the source map binning and orientation.}
	\item{Coregister and add residual maps together to produce the total residual map.}
\end{enumerate}

Example residual maps, along with the corresponding observations and source map, are shown in Figure \ref{fig:resid}. Both the raw \textit{FOXSI-2} images and individual residual maps are plotted in the corresponding detector plane, with each detector rotated at a different angle relative to solar north. After 5 iterations of the deconvolution procedure, there are no distinct features in the residual maps, indicating that the source represents the observations reasonably well. In the total residual map and combined (source + residuals) map, we observe artifacts of the coarse \textit{FOXSI-2} strip crossings, which are much larger than the source map pixels (${\sim}1"$ in the example). With the combined map, we can directly compare the \textit{FOXSI-2} images to \textit{RHESSI} CLEAN images and asses the imaging dynamic ranges (see Figure \ref{fig:rhessi_intensity}).\\


\begin{thebibliography}{}

\bibitem[Aschwanden et al.(2016)]{aschw2016} Aschwanden, M.~J., Holman, G., O'Flannagain, A., et al.\ 2016, \apj, 832, 27

\bibitem[Athiray et al.(2017)]{athiray2017} Athiray, P.~S., Buitrago-Casas, J.~C., Bergstedt, K., et al.\ 2017, \procspie, 103970A

\bibitem[Athiray et al.(2020)]{athiray2020} Athiray, P.~S., Vievering, J., Glesener, L., et al.\ 2020, \apj, 891, 1 

\bibitem[Battaglia et al.(2005)]{battaglia2005} Battaglia, M., Grigis, P.~C., \& Benz, A.~O.\ 2005, \aap, 439, 737 

\bibitem[Benvenuto et al.(2013)]{benvenuto2013} Benvenuto, F., Schwartz, R., Piana, M., et al.\ 2013, \aap, 555, A61

\bibitem[Boerner et al.(2012)]{boerner2012} Boerner, P., Edwards, C., Lemen, J., et al.\ 2012, \solphys, 275, 41

\bibitem[Brown(1971)]{brown1971} Brown, J.~C.\ 1971, \solphys, 18, 489 

\bibitem[Buitrago-Casas et al.(2017)]{milo2017} Buitrago-Casas, J.~C., Elsner, R., Glesener, L., et al.\ 2017, \procspie, 10399, 103990J

\bibitem[Christe et al.(2008)]{christe2008} Christe, S.~D., Hannah, I., Krucker, S., et al.\ 2008, AGU Spring Meeting Abstracts, SP51C-09 

\bibitem[Christe et al.(2016)]{christe2016} Christe, S., Glesener, L., Buitrago-Casas, C., et al.\ 2016, Journal of Astronomical Instrumentation , 5, 1640005-625 

\bibitem[Cooper et al.(2020)]{cooper2020} Cooper, K., Hannah, I.~G., Grefenstette, B.~W., et al.\ 2020, \apjl, 893, L40

\bibitem[Del Zanna(2013)]{delzanna2013} Del Zanna, G.\ 2013, \aap, 558, A73


\bibitem[Emslie et al.(2012)]{emslie2012} Emslie, A.~G., Dennis, B.~R., Shih, A.~Y., et al.\ 2012, \apj, 759, 71 

\bibitem[Glesener et al.(2016)]{glesener2016} Glesener, L., Krucker, S., Christe, S., et al.\ 2016, \procspie, 99050E

\bibitem[Glesener et al.(2017)]{glesener2017} Glesener, L., Krucker, S., Hannah, I.~G., et al.\ 2017, \apj, 845, 122 

\bibitem[Glesener et al.(2020)]{glesener2020} Glesener, L., Krucker, S., Duncan, J., et al.\ 2020, \apjl, 891, L34

\bibitem[Grefenstette et al.(2016)]{grefen2016} Grefenstette, B.~W., Glesener, L., Krucker, S., et al.\ 2016, \apj, 826, 20 

\bibitem[Hannah et al.(2008)]{hannah2008} Hannah, I.~G., Christe, S., Krucker, S., et al.\ 2008, \apj, 677, 704 

\bibitem[Hannah et al.(2016)]{hannah2016} Hannah, I.~G., Grefenstette, B.~W., Smith, D.~M., et al.\ 2016, \apjl, 820, L14

\bibitem[Hannah et al.(2019)]{hannah2019} Hannah, I.~G., Kleint, L., Krucker, S., et al.\ 2019, \apj, 881, 109

\bibitem[Harrison et al.(2013)]{harrison2013} Harrison, F.~A., Craig, W.~W., Christensen, F.~E., et al.\ 2013, \apj, 770, 103 

\bibitem[Hudson(1991)]{hudson1991} Hudson, H.~S.\ 1991, \solphys, 133, 357

\bibitem[Ishikawa et al.(2011)]{ishikawa2011} Ishikawa, S., Saito, S., Tajima, H., et al.\ 2011, IEEE Transactions on Nuclear Science, 58, 2039

\bibitem[Ishikawa et al.(2016)]{ishikawa2016} Ishikawa, S.-n., Katsuragawa, M., Watanabe, S., et al.\ 2016, Journal of Geophysical Research (Space Physics), 121, 6009 

\bibitem[Ishikawa et al.(2017)]{ishikawa2017} Ishikawa, S.-. nosuke ., Glesener, L., Krucker, S., et al.\ 2017, Nature Astronomy, 1, 771

\bibitem[Isola et al.(2007)]{isola2007} Isola, C., Favata, F., Micela, G., \& Hudson, H.~S.\ 2007, \aap, 472, 261 

\bibitem[Klimchuk(2006)]{klimchuk2006} Klimchuk, J.~A.\ 2006, \solphys, 234, 41

\bibitem[Kontar et al.(2019)]{kontar2019} Kontar, E.~P., Jeffrey, N.~L.~S., \& Emslie, A.~G.\ 2019, \apj, 871, 225

\bibitem[Krucker et al.(2008)]{krucker2008} Krucker, S., Battaglia, M., Cargill, P.~J., et al.\ 2008, \aapr, 16, 155

\bibitem[Krucker et al.(2013)]{krucker2013} Krucker, S., Christe, S., Glesener, L., et al.\ 2013, \procspie, 88620R

\bibitem[Krucker et al.(2014)]{krucker2014} Krucker, S., Christe, S., Glesener, L., et al.\ 2014, \apjl, 793, L32 

\bibitem[Kuhar et al.(2018)]{kuhar2018} Kuhar, M., Krucker, S., Glesener, L., et al.\ 2018, \apjl, 856, L32 

\bibitem[Lemen et al.(2012)]{lemen2012} Lemen, J.~R., Title, A.~M., Akin, D.~J., et al.\ 2012, \solphys, 275, 17

\bibitem[Lin et al.(2002)]{lin2002} Lin, R.~P., Dennis, B.~R., Hurford, G.~J., et al.\ 2002, \solphys, 210, 3 

\bibitem[Lucy(1974)]{lucy1974} Lucy, L.~B.\ 1974, \aj, 79, 745

\bibitem[Neupert(1968)]{neupert1968} Neupert, W.~M.\ 1968, \apjl, 153, L59

\bibitem[Newville et al.(2014)]{newville2014} Newville, M., Stensitzki, T., Allen, D.~B., \& Ingargiola, A.\ 2014, LMFIT: Non-Linear Least-Square Minimization and Curve-Fitting for Python, 1.0.0, Zenodo

\bibitem[Parker(1988)]{parker1988} Parker, E.~N.\ 1988, \apj, 330, 474 

\bibitem[Peterson \& Cote(1997)]{peterson1997} Peterson, G.~L. \& Cote, M.\ 1997, \procspie, 3113, 321

\bibitem[Richardson(1972)]{richardson1972} Richardson, W.~H.\ 1972, Journal of the Optical Society of America (1917-1983), 62, 55 

\bibitem[Wright et al.(2017)]{wright2017} Wright, P.~J., Hannah, I.~G., Grefenstette, B.~W., et al.\ 2017, \apj, 844, 132 

\end{thebibliography}
\end{document}